\documentclass[11pt,preprint,eqsecnum]{aastex}
\usepackage{graphicx,rotating,amsmath,amssymb,epsfig,natbib,float,threeparttable}
\slugcomment{Accepted to ApJ}
\shorttitle{Zombie Vortices in PPD}
\shortauthors{MPJBHL2015}

\newcommand{\bfome}{\mbox{\boldmath $\omega$}}
\newcommand{\bfDel}{\mbox{\boldmath $\nabla$}}

\newcommand*{\eg}{e.g.,\ }
\newcommand*{\ie}{i.e.,\ }
\newcommand*{\cf}{cf.,\ }

\begin{document}

\title{Zombie Vortex Instability I: A Purely Hydrodynamic Instability to Resurrect the Dead Zones of Protoplanetary Disks}

\author{Philip S. Marcus, Suyang Pei, Chung-Hsiang Jiang}
\affil{Department of Mechanical Engineering, University of California, Berkeley}
\affil{6121 Etcheverry Hall, Mailstop 1740, Berkeley, CA 94720-1740}
\email{pmarcus@me.berkeley.edu}

\author{Joseph A. Barranco}
\affil{Department of Physics \& Astronomy, San Francisco State University}
\affil{1600 Holloway Avenue, San Francisco, CA 94132}

\author{Pedram Hassanzadeh}
\affil{Center for the Environment and Department of Earth \& Planetary Sciences, Harvard University}
\affil{20 Oxford Street, Cambridge, MA 02138}

\and

\author{Daniel Lecoanet}
\affil{Department of Astronomy, University of California, Berkeley}
\affil{501 Campbell Hall, Berkeley, CA 94720}

\begin{abstract}
There is considerable interest in hydrodynamic instabilities in dead zones of protoplanetary disks as a mechanism for driving angular momentum transport and as a source of particle-trapping vortices to mix chondrules and incubate planetesimal formation.  We present simulations with a pseudo-spectral anelastic code and with the compressible code {\it Athena}, showing that {\it stably stratified} flows in a shearing, rotating box are violently unstable and produce space-filling, sustained turbulence dominated by large vortices with Rossby numbers of order $\sim$$0.2-0.3$.  This {\it Zombie Vortex Instability} (ZVI) is observed in both codes and is triggered by Kolmogorov turbulence with Mach numbers less than $\sim$$0.01$.  It is a common view that if a given constant density flow is stable, then stable vertical stratification should make the flow even more stable.  Yet, we show that sufficient vertical stratification can be unstable to ZVI.  ZVI is robust and requires no special tuning of boundary conditions, or initial radial entropy or vortensity gradients  (though we have studied ZVI only in the limit of infinite cooling time).  The resolution of this paradox is that stable stratification allows for a new avenue to instability: baroclinic critical layers.  ZVI has not been seen in previous studies of flows in rotating, shearing boxes because those calculations frequently lacked vertical density stratification and/or sufficient numerical resolution.  Although we do not expect appreciable angular momentum transport from ZVI in the small domains in this study, we hypothesize that ZVI in larger domains with compressible equations may lead to angular transport via spiral density waves.
\end{abstract}


\section{INTRODUCTION}

\subsection{Can Non-Magnetically-Coupled, Non-Self-Gravitating Keplerian Disks Have Purely Hydrodynamic Instabilities? Can They Drive Angular Momentum Transport?}

After 40 years of intense theoretical and computational research, the answer has been a qualified -- and unsatisfying -- ``maybe.''  It has long been known that molecular viscous diffusion is wholly inadequate to drive accretion because the transport timescale is several orders of magnitude longer than the age of the universe: $\tau_{visc}\sim R^2/\nu\sim 100,000$~Gyr for conditions typical at $R\sim 1$~AU in a protoplanetary disk (PPD) with kinematic viscosity $\nu\sim v_{th}\ell$, thermal velocity $v_{th}\sim 1$~km/s, molecular mean free path $\ell\sim$ 1~cm. \citet{SS73} proposed that Keplerian differential rotation may be susceptible to hydrodynamic and/or magnetohydrodynamic instabilities that would drive turbulent diffusion, for which they introduced an ``eddy'' viscosity: $\nu_{eddy}=\alpha c_sH$ (the idea being that turbulent ``blobs'' would move at subsonic speeds over distances of order the scale height, and the parameter $\alpha$ would subsume all our ignorance over the details of the transport efficiency).  Because the specific angular momentum increases with radius in Keplerian differential rotation, it is often stated that Keplerian shear is stable to purely hydrodynamic instabilities. However, this centrifugal stability criterion \citep{R17,synge1933} is strictly true only for axisymmetric perturbations in an unstratified, inviscid flow, and therefore is of limited relevance for {\it stratified} astrophysical disks. 
 
So, what is the nature of hydrodynamic instabilities that may lead to enhanced transport?  Historically, by analogy with Poiseuille (pressure-driven) and Couette (viscous-driven) flows in both planar and cylindrical geometries, the go-to mechanism has been subcritical finite-amplitude instabilities that extract energy from the shear at very high Reynolds numbers \citep{richard1999}.  This has been, and remains, highly controversial.  \citet{balbus96} and \citet{hawley99} have argued, with theory and computational analyses, that the Coriolis force stabilizes Keplerian shear and shuts off any way (in the absence of magnetic fields) for perturbations to extract energy from the differential rotation in order to power sustained turbulence.  Others have argued that even the best to-date numerical simulations lack sufficient resolution to capture nonlinear triggering of instabilities at very high Reynolds numbers \citep{longaretti2002, richard2003, dubrulle2005, lesur2005}.  A related mechanism to generate shear turbulence starts with the transient amplification of a special class of initial conditions, which may trigger nonlinear feedback and continued growth if the amplitude of the transients peak above some critical threshold \citep{chagelishvili2003, tevzadze2003, umurhan2005, afshordi2005, mukhopadhyay2005, tevzadze2008,salhi2010,salhi2013}.  However, none of these ``by-pass to turbulence" approaches regenerate the initial perturbations required to sustain turbulence after the initial burst of turbulence has decayed.

With conflicting results in both theory and numerical work, one may hope that laboratory experiments may provide an independent test.  Unfortunately, two recent Taylor-Couette flow experiments (water rotating between two concentric cylinders) have yielded opposite results: experiments by \citet{PL11} claim a positive measurements of turbulence and outward transport of angular momentum; whereas, a similar set of  experiments by  \citet{ji06,schartman12} claim no such evidence of turbulence nor angular momentum transport.  The measurements are  difficult and sensitive to the ability of the experimentalists to control secondary flows from their apparatus (\eg Ekman pumping).  Direct numerical simulations of Taylor-Couette flows with laboratory boundary conditions seem to confirm that the axial boundaries are driving the instabilities and enhanced transport \citep{avila2012}.

\citet{velikhov59} and \citet{chandra60,chandra61} showed that magnetic fields could destabilize otherwise stable rotational flows.  Balbus and Hawley were the first to apply this magnetorotational instability (MRI) to Keplerian shear flows and demonstrate how the resulting turbulence efficiently transported angular momentum outward \citep{balbus91,hawley91}.  However, there exist relatively dense, cool and nearly neutral regions in PPDs (approx.\ 1-10 AU) that likely lack sufficient coupling between matter and magnetic fields \citep{blaes94}, except perhaps in thin surface layers that have been ionized by cosmic rays or protostellar X-rays \citep{gammie96}.  The magnetically-decoupled regions are known as ``dead zones'' and there is continued research into whether waves and turbulence from the magnetically-coupled MRI-active regions can propagate into the dead zones of PPDs and drive motions that result in angular momentum transport or disrupt the settling of dust \citep{FS03,turner2007,turner2008,OML09}.

In regions of a laminar near-Keplerian disk that are strongly coupled to magnetic fields, there is no question that of all the known relevant linear instabilities, MRI is the fastest to grow to large amplitude.  However, the presence of dead zones in PPDs still motivates researchers to investigate other mechanisms for instability and transport, especially because the dead zone seems to be coincident with planet-forming regions within disks \citep{matsumura2009,matsumura2007}.  A large class of potential instabilities involve global properties of the disk, specifically radial gradients of thermodynamic quantities.  Extrema in the radial profile of potential vorticity (a.k.a.\ vortensity) is unstable to Rossby waves which eventually roll-up into vortices \citep{lovelace99,li00,li01,varniere2006,richard2013,meheut2010,meheut2012a,meheut2012b}.  Radial entropy gradients that are stable according to the Solberg-H\o iland criterion can still be unstable to a ``subcritical baroclinic instability'' (SBI) that is most effective when the thermal cooling time is of order an orbital period \citep{klahr03, petersen2007a,petersen2007b,lesur2010,lyra2011}.  More recent work on the SBI by \citet{klahr2014} has shown that the instability is actually linear (\ie does not require large amplitude perturbations) and more akin to a convective overstability rather than a traditional baroclinic instability.  Still others have investigated vertical shear instabilities (VSI) related to the Goldreich-Shubert-Fricke instability in rotating stars \citep{nelson2013,goldreich1967,fricke1968}.  While ZVI needs vertical stable stratification (\eg entropy increasing in direction opposite of gravity and sufficiently long thermal relaxation times) the VSI needs vertically neutral stratification (\eg no vertical entropy gradient, or sufficiently short thermal relaxation times). Thus depending on the thermodynamic structure of the disk and its optical properties, one or the other might locally dominate.

Of course, by direct observation, we know that {\it some} mechanisms exist within a PPD that leads to the transport of angular momentum in the way that is needed to complete star formation and that allows dust to accumulate and agglomerate into kilometer-sized planetesimals.  However, the elucidation of that mechanism or mechanisms cannot be sloughed off as an unimportant ``detail.''   A more realistic treatment of turbulence is essential to quantify the spatial and temporal distribution of dust in a PPD and therefore affects our modeling of the infrared emission from dust (relevant for ALMA \& JWST observations), as well as an initial-mass function (IMF) for planets.

\subsection{A New Instability That Leads to Space-Filling Robust ``Zombie'' Vortices \& Turbulence}\label{sec:cartoon}

In \citet[][hereafter BM05]{barranco05}, we investigated the stability of three-dimensional vortices in the midplanes of protoplanetary disks.  We created a model for a near-equilibrium vortex, which we initialized by hand in our spectral anelastic code that had specially tailored algorithms to handle shear, rotation and stratification \citep{barranco06}.  The vortex suffered a linear instability which resulted in its destruction near the midplane, but we unexpectedly discovered that the stratified regions away from the midplane naturally filled with robust vortices.  At the time, we speculated that the initial midplane vortex excited internal gravity waves, which propagated away from the midplane and deposited their energy in stratified regions where the shear rolled vorticity perturbations into new vortices.

However, our original explanation was not entirely complete nor satisfying.  \citet[][hereafter MPJH13]{MPJH13} correctly diagnosed the true mechanism for the formation of these vortices and in the process identified a new instability, which we now call the ``Zombie Vortex Instability" or ZVI.   In order to get at the essential nature of the phenomenon, MPJH13 stripped out complicating features of a protoplanetary disk (\eg spatially varying gravity and Brunt-V\"{a}is\"{a}l\"{a} frequency) and numerically investigated simple Couette flow with constant gravity and constant Brunt-V\"{a}is\"{a}l\"{a} frequency in the limit of the Boussinesq approximation.  In this far simpler system, MPJH13 found that a small perturbing vortex could trigger an instability in a rapidly-rotating, strongly stratified flow, yielding vigorous, space-filling vortices, vortex layers and turbulence.

\begin{figure}
\epsscale{1.0}
\plotone{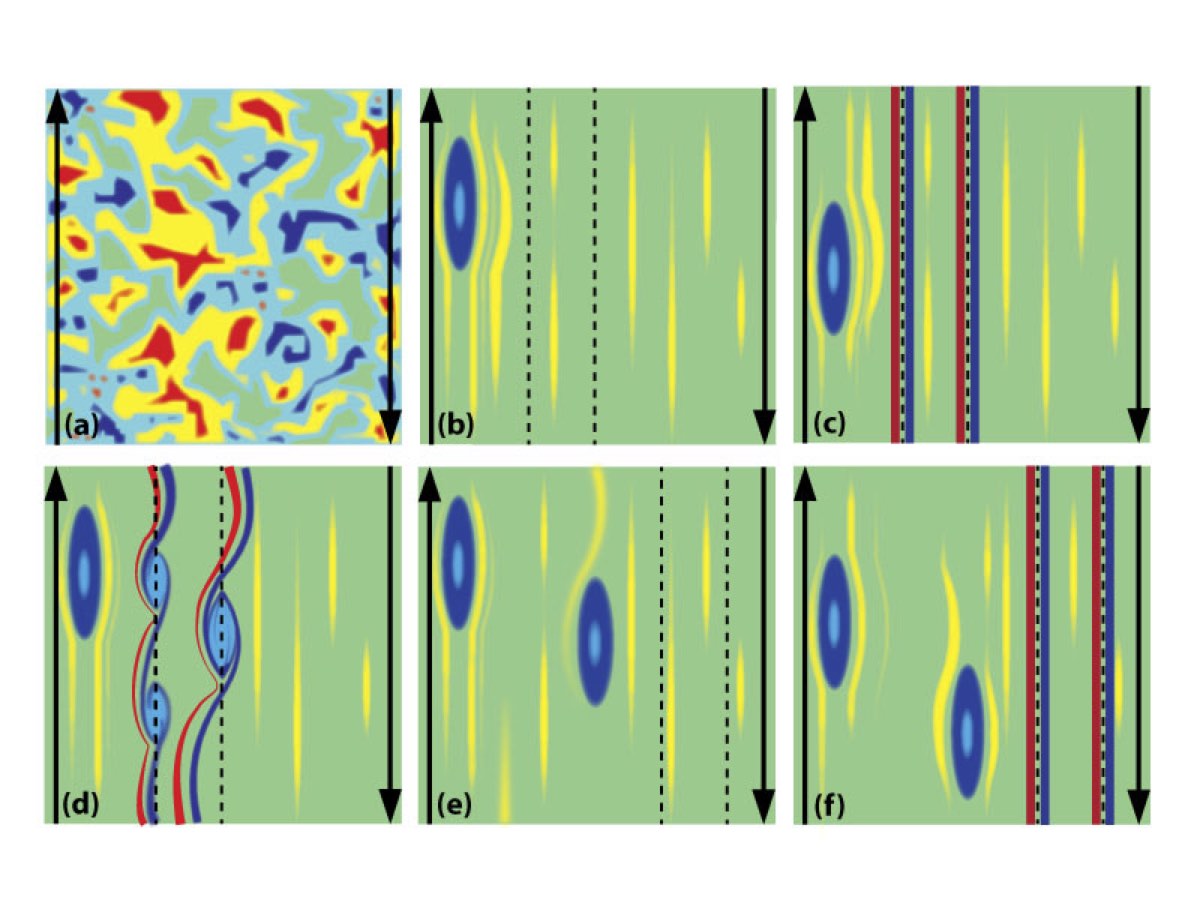}
\caption{\label{F:zombie_cartoon} Simplified cartoon of zombie vortex instability. See \S 1.2 for a more detailed explanation.  Each panel shows a small patch of a PPD in frame rotating with the gas; upward is the forward azimuthal direction, rightward is increasing radius from protostar.  In the rotating frame, the background flow appears as an anticyclonic shear (indicated with the arrows). Color illustrates vertical vorticity (aligned with the rotation axis of the disk) relative to the background Keplerian shear, with red indicating cyclonic (out of the page) vorticity and blue indicating anticyclonic (into the page) vorticity. (a) Domain is initially filled with Kolmogorov turbulence.  In time, the background shear stretches cyclonic regions into vortex layers, while anticyclonic regions roll-up into compact vortices.  Smaller vortices merge into larger vortices, a feature of ``reverse cascades'' in rotating shear flows.  (b) After a period of vortex mergers, a small region will have a few  isolated anticyclones.  Here, we show one dominant anticyclone.  The dotted lines indicates the location of the $m=1$ and $m=2$ critical layers. (c) Anticyclone acts as a source of forcing, exciting neighboring critical layers and creating dipolar vortex layers.  
(d) Vortex layers with same orientation as background shear are linearly unstable.  (e)  Anticyclonic vortex layers roll-up into compact anticyclones, which tend to merge into one larger vortex.  (f) New vortex excites neighboring critical layers and the process repeats. The instability is inherently 3D. Although the schematic shows vortices and critical layers forming only on the right side of the surviving vortex in (b), the critical layers and vortices form on the left side as well. }
\end{figure}

Figure \ref{F:zombie_cartoon} is a cartoon picture summary of the mechanism outlined in MPJH13.  Each panel shows a small patch of a PPD in frame rotating with the gas; upward is the forward azimuthal direction, rightward is increasing radius from the protostar.  In the rotating frame, the background flow appears as an anticyclonic shear (as indicated with the arrows). Color illustrates vertical vorticity (aligned with the rotation axis of the disk), with red indicating cyclonic (out of the page) vorticity and blue indicating anticyclonic (into the page) vorticity.  The domain is initially filled with Kolmogorov turbulence on small scales (panel a).  At low Rossby number (when the Coriolis effect is dominant over advection), the turbulence can undergo a ``reverse cascade'' in which energy transfers to larger scales as small regions of vorticity merge with other regions of vorticity with the same sense of rotation.  Regions of vorticity that have the same sense of rotation as the background shear tend to roll-up into compact vortices, whereas vorticity patches with the opposite sense of rotation as the background shear tend to get stretched out into thin layers \citep{marcus93}.  After a period of reverse cascades and mergers, a region of the disk may only be populated with a few isolated anticyclones that move with the shear and interspersed with cyclonic vortex layers (panel b).  In our simplified cartoon, we consider a small region with only one remaining anticyclone.   This vortex acts as a source of perturbations and waves.

The crucial new insight in MPJH13 was the recognition of ``baroclinic critical layers'' as being sites that are receptive to perturbations.  Critical layers are special locations in a shear flow where the coefficients of the highest derivatives of the linearized equations vanish, indicating that the neutrally-stable eigenmodes are singular there \citep{maslowe86, drazin81}.  Because our focus in this paper is to demonstrate the existence of ZVI, we will defer the mathematics of critical layers to a future paper.  For our purposes here, we note that there exist multiple critical layers on either side of the source vortex.  Let $(x,y,z)$ be local Cartesianized radial, azimuthal, and vertical directions within a small box rotating with the gas.  The cross-stream distance $\delta(m)$ between the source and the $m$th critical layer is given by (MPJH13):
\begin{equation}\label{delta}
\delta(m)\equiv\frac{N}{|\sigma|}\frac{1}{k_y}=\frac{N}{|\sigma|}\frac{L_y/2\pi}{m}\equiv\frac{\Delta}{m},
\end{equation}
where $N$ is the Brunt-V\"{a}is\"{a}l\"{a} frequency, $\sigma$ is the local shear rate, $k_y$ is a spatial wavenumber associated with a Fourier component of the source vortex, $L_y$ is the domain size in the azimuthal direction, and $m$ is some non-zero integer.   The critical layer closest to the source, without intersecting the source itself, has the largest integer $m_{max}$.  At first glance, it may appear that the closest critical layer depends on the arbitrary choice of the domain size in the azimuthal direction, but in numerical experiments, the physical location of the first excited critical layer is always just outside the perturber, and the value of the integer $m_{max}$ will scale with $L_y$. Critical layers for smaller integers $m \le m_{max}$ are also excited. The critical later farthest from the source generally has $m=1$ and is at a distance $\delta(1) \equiv \Delta$. However, if a boundary of the domain in the cross-stream direction is too close to the source, or if there are  physics that prohibit vortex coherence over distances as large $\Delta$, then the excited critical layer farthest from the perturbing source will have $m>1$. In the calculations presented here the boundaries are chosen so that the farthest excited critical layer is 1, and the equations of motion contain no physics that inhibit large vortices from forming, so the farthest distance is $\Delta$. However, as we discuss \S \ref{subsec:mach}, the physics of the compressible gas in PPDs will generally limit the size of vortices such that the farthest critical layers do not have $m=1$.

While the singular eigenmodes are neutrally stable (do not grow nor decay) to small perturbations, they are susceptible to being forced and excited by the nearby anticyclone.  It is important to note that although the cartoon makes it appear that the excited critical layers are at the same height as the perturber, they are in fact located somewhat above and below -- the instability is inherently 3D.  A dipolar vortex layer (two juxtaposed oppositely-signed layers of vorticity) develops at the location of each critical layers (panel c).  Vortex layers that have the same sense of rotation as the background shear are linearly unstable (panel d) and roll-up into compact vortices \citep{marcus93}, and so new anticyclones will eventually be spawned at the locations of the critical layers with $m$ vortices at the $m^{th}$ critical layer (panel e).  The newly spawned vortices from the different critical layers tend to merge into one larger vortex with that vortex located at 
the critical layer location farthest from the source (generally with $m=1$ at a distance $\Delta$). The newly created large vortex then excites its neighboring critical layers (panel f) and the process continues.  In numerical simulations, one can observe perturbations spawning new vortices farther and farther from the original perturbation vortex.

We named the instability the Zombie Vortex Instability not only because it occurs in the dead zones of PPD, but also because of the way one zombie vortex ``infects'' its neighboring critical layer, spawning new zombie vortices, which ``infect'' their neighboring critical layers, etc.  In numerical simulations, one can see the entire domain rapidly fill with zombie vortices from one initial vortex.  The instability is not an artifact of the numerical method as we have observed it with spectral codes and finite-volume codes (\eg {\it Athena}), with fully compressible, anelastic and Boussinesq treatments of the continuity condition, with and without the shearing box, and with either hyperviscosity or real molecular viscosity.  One may ask, if it is so robust, how was it missed in previous numerical calculations?  As will be elucidated in this series of papers, ZVI requires vertical stratification, high resolution to resolve the narrow critical layers, a broad spectrum of perturbations (\ie Kolmogorov, but not Gaussian-peaked), and enough simulation time to allow the critical layers to amplify perturbations.

\subsection{Paper Outline}

In this paper, we consider ZVI in the astrophysical context, rather than the Boussinesq approximation appropriate to laboratory Couette flow as we did in MPJH13.  In \S \ref{sec:equations}, we review the hydrodynamic equations of motion and outline the parameter space of interest for ZVI. Because the primary tuning parameter for ZVI is the Brunt-V\"{a}is\"{a}l\"{a} frequency $N(z)$, we shall replace linear gravity with a spatially constant value so that $N$ itself is also constant (for vertically isothermal backgrounds). Our previous numerical experiments (including those in BM05) show that ZVI is present even when  $N(z)$ is spatially varying.  However, those experiments also convinced us that dealing with a constant $N$ makes it much easier to interpret the results and provide a pedagogical explanation for them.  In \S \ref{sec:evidence}, we shall study ZVI and its development into turbulence.  We shall also investigate triggers of ZVI other than an initial vortex (which was used as the initial condition in BM05 and MPJH13), namely initial isotropic, homogeneous Kolmogorov-like noise.  We shall characterize properties of zombie turbulence: its magnitude, its space-filling nature, and its inhomogeneity and anisotropy due to persistent, but turbulent, large anticyclonic vortices and undulating cyclonic vortex layers. 

Most of our numerical experiments presented here use the anelastic approximation (which filters acoustic waves) with our pseudo-spectral code that has specially tailored algorithms to handle shear, rotation and stratification \citep{barranco06}.  We have also done simulations with the fully compressible, Godunov finite-volume code {\it Athena} \citep{gs08,stone08,sg10}, and show that qualitatively and quantitatively we observe similar results.  However, our timing tests show that {\it Athena} is approximately 100 times more computationally expensive for these numerical experiments when using the same number of grid points as spectral modes in the pseudo-spectral code.

In our discussion in \S \ref{sec:discussion}, we explain why ZVI was not observed in previous studies by others, we review ZVI in the context of other hydrodynamic and magnetohydrodynamic instabilities in PPDs, and we consider the role of ZVI in angular momentum transport and planetesimal formation.  In a future paper, we will confront ZVI on a more mathematical and theoretical level and address thresholds required to trigger ZVI. In particular, we will show why extraordinarily weak Kolmogorov-like noise is so efficient in exciting ZVI. 

\section{HYDRODYNAMIC EQUATIONS IN A CARTESIAN DOMAIN} \label{sec:equations}

Consider a PPD in which the steady flow is only in the azimuthal direction with angular velocity
\begin{equation}
\Omega(R) \propto R^{-q}, \label{q}
\end{equation}
where $R$ is the cylindrical radius from the protostar.  A value of $q=1.5$ corresponds to Keplerian rotation.  Now consider a three-dimensional box of gas centered at radius $R_0$ from the protostar that co-rotates with the gas with angular rate $\Omega_0\equiv\Omega(R_0)$.  The box is sufficiently small that we can ignore curvature of the disk and map the cylindrical coordinates onto local Cartesian coordinates: $(R-R_0,R_0\phi,z)\longrightarrow(x,y,z)$ \citep{hill1878,goldreich65b}.  In what follows, symbols with ``hats'' will denote unit vectors: $\hat{\bf x}$ points in the local outward radial direction, $\hat{\bf y}$ points in the local azimuthal direction, and $\hat{\bf z}$ points in the local vertical direction.

\subsection{Compressible Euler Equations}

Euler's equations (continuity, momentum and energy with no viscous, radiative or diffusive terms) for the gas in the rotating frame are:
\begin{mathletters}\label{BHS35}
\begin{eqnarray}
{\frac{\partial \rho}{\partial t}} + \bfDel \cdot (\rho {\bf v}) &=& 0, \label{BHS35a} \\
{\frac{\partial {\bf v}}{\partial t}}+ ({\bf v} \cdot \bfDel){\bf v} &=& -{\frac{1}{\rho}}\bfDel{P} -g(z)\hat{\bf z}-2 \Omega_0 \, \hat{\bf z} \times {\bf v} +2q \Omega_0^2 \, x \, \hat{\bf x}, \label{BHS35b} \\
{\frac{\partial\epsilon}{\partial t}} + ({\bf v} \cdot \bfDel)\epsilon &=& - \frac{P}{\rho}(\bfDel \cdot {{\bf v}}), \label{BHS35c}
\end{eqnarray}
\end{mathletters}
\noindent where ${\bf v}(x, y, z,t)$ is the gas velocity with respect to the center of the box that co-rotates with the PPD.  $P$, $\rho$, and $\epsilon$ are, respectively, the gas pressure, density and specific internal (thermal) energy.  The vertical acceleration of gravity $-g(z)$ does not include any contribution from the self-gravity of the gas itself.  The term $-2 \Omega_0 \, \hat{\bf z} \times {\bf v}$ is the Coriolis force, and the term $2q\Omega_0^2 \, x\hat{\bf x}$ is the tidal acceleration that arises from the difference between the inward central force that causes circular motion $R\Omega^2(R)$ and the outward centrifugal force $R\Omega_0^2$ due to being in a rotating frame.  The term on the right-hand side of the thermal energy equation is the rate at which pressure forces do work due to volume changes of fluid parcels.  To close the system of equations, we assume an ideal gas for the equation of state:
\begin{equation}
P = (\gamma-1) \rho \epsilon = \mathcal{R} \rho T, \label{BHS35d}
\end{equation}
where $T$ is the gas temperature, $\mathcal{R}$ is the gas constant, and $\gamma\equiv C_P/C_V$ is the adiabatic index, or ratio of specific heats at constant pressure and constant volume.  Throughout this paper, we set ratio of specific heats $\gamma=5/3$.

A time-independent equilibrium (denoted with overbars) to the above set of equations is:
\begin{mathletters} \label{hydro}
\begin{eqnarray}
{\bf \bar{v}} &=& \left(\bar{v}_x,  \bar{v}_y, \bar{v}_z\right) = \left(0, -q \Omega_0 \,  x, 0\right), \label{hydrostatic2} \\
\frac{d\bar{P}(z)}{dz} &=& - \bar{\rho}(z) g(z). \label{hydrostatic}
\end{eqnarray}
\end{mathletters}
Because there is no  radiative or diffusive effects in equations~(\ref{BHS35}), there is a degeneracy of allowable equilibrium thermodynamic profiles, so
the equilibrium temperature $\bar{T}(z)$ corresponding to the equilibrium velocity $\bar{\bf v}$ is not unique.  In fact, one thermodynamic quantity, $\bar{P}(z)$,  $\bar{T}(z)$,  $\bar{\rho}(z)$,  or $\bar{\epsilon}(z)$ can be arbitrarily specified, and the others will follow.  In this work, we shall assume $\bar{T}(z)=T_0=~constant$, so:
\begin{equation}
\frac{\bar{P}(z)}{\mathcal{R}T_0} = \bar{\rho}(z) = \rho_0\exp\left[-\int_0^z\frac{g(z')}{\mathcal{R}T_0}~dz'\right], \label{hydrostatic1}
\end{equation}
where $\rho_0$ is the midplane density and $z=0$ is the midplane of the disk.

\subsection{Anelastic Approximation}

The anelastic approximation has been extensively used in the study of deep, subsonic convection in planetary atmospheres \citep{ogura62,gough69,bannon96} and stars 
\citep{gilman81,glatzmaier81a,glatzmaier81b}. We have previously used the anelastic approximation to study three-dimensional vortices in PPDs \citep{barranco00a,barranco05,barranco06} and the Kelvin-Helmholtz instability of settled dust layers in PPDs \citep{barranco09,lee2010a,lee2010b}.  The basic idea is that there may be large variations in the background pressure and density in hydrostatic equilibrium, but that at any height in the atmosphere, the fluctuations of the pressure and density are small compared to the background values at that height.  Mathematically,  all flow variables are expanded in powers of the Mach number (presumed to be small), and only lowest order terms are retained in the equations of motion.  In general, the anelastic approximation results in a valid description of the flow when:
\begin{mathletters}\label{E:anelastic_validity}
\begin{eqnarray}
\frac{|\mathbf{v}(x,y,z,t)|}{c_s} &\equiv&Ma \ll 1,\\
\frac{|\rho(x,y,z,t)-\bar{\rho}(z)|}{\bar{\rho}(z)} \sim \frac{|P(x,y,z,t)-\bar{P}(z)|}{\bar{P}(z)} &\sim& Ma^2 \ll 1,
\end{eqnarray}
\end{mathletters}
where $c_s$ is the local sound speed.  Diagnostically, it is trivial to check every step during a numerical calculation whether the conditions (\ref{E:anelastic_validity}) are satisfied.  

There are a number of variations of the anelastic approximation, all equivalent to the same order of the Mach number, but varying in whether they retain or drop certain smaller terms or in exactly how they linearize the equation of state, often with an eye toward conserving some quantity such as energy; see references above as well as \citet{brown2012,vasil2013}.   For the special case of uniform temperature equilibrium background, the most straight-forward application of the anelastic approximation yields energy-conserving equations, so we shall defer going into more detail until future papers in which we consider different background temperature profiles.  The Euler equations with the anelastic approximation for a vertically uniform background temperature are:
\begin{mathletters}\label{E:anelastic_eqns}
\begin{eqnarray}
\bfDel \cdot [\bar{\rho}(z) {\bf v}] &=& 0, \label{anelastic1}\\
{\frac{\partial {\bf v}}{\partial t}}  &=& -({\bf v}\cdot \bfDel){\bf v}-2\Omega_0 \, \hat{\bf z} \times {\bf v} +2q \Omega_0^2 \, x \, \hat{\bf x}\nonumber\\
&{}&-\bfDel\left[\frac{{P - \bar{P}(z)}}{{\bar{\rho}(z)}}\right] + \left[{\frac{T - T_0}{T_0}}\right] g(z) \, \hat{\bf z}, \label{E:anelastic2}\\
{\frac{\partial T}{\partial t}} &=& - ({\bf v} \cdot \bfDel) T  - \left[\frac{N^2(z)}{g(z)}\right]Tv_z, \label{E:anelastic3}\\
\left[{\frac{P - \bar{P}(z)}{\bar{P}(z)}}\right] &=&\left[{\frac{\rho - \bar{\rho}(z)}{\bar{\rho}(z)}}\right] + \left[{\frac{T - T_0}{T_0}}\right], \label{E:anelastic4}
\end{eqnarray}
\end{mathletters}
where $N(z)$ is the Brunt-V\"{a}is\"{a}l\"{a} frequency, a measure of the convective stability of the atmosphere.  In general, $N^2\propto \vec{g}\cdot\bfDel\bar{s}$, where $\bar{s}$ is the background entropy.  Here, we focus on vertical entropy gradients, so:
\begin{mathletters}
\begin{eqnarray}
N^2(z)&\equiv& (g/C_P)(d\bar{s}/dz),\\
{}&=& g^2/(C_PT_0)\quad\quad\mathrm{(for~an~isothermal~hydrostatic~background)}. \label{more}
\end{eqnarray}
\end{mathletters}
One of the main dynamical consequences of this approximation is that the total density is replaced by the time-independent mean density in the mass continuity equation, which has the effect of filtering acoustic waves, but allowing slower wave phenomena such as internal gravity waves, Rossby waves, etc.  

Similarly to the fully compressible equations, the dissipationless anelastic equations contain a degeneracy, so any one thermodynamic quantity, $\bar{P}(z)$,  $\bar{T}(z)$,  $\bar{\rho}(z)$,  or $\bar{\epsilon}(z)$ can be arbitrarily specified.  As before, we arbitrarily set the background equilibrium temperature to be constant, $\bar{T}(z)=T_0$.  We note that the isothermal equilibrium solutions to the fully compressible equations are the same isothermal equilibrium solutions to the anelastic equations.

\subsection{Background Equilibrium States}

In a PPD, the vertical component of the protostellar gravity varies linearly with distance from the midplane, which yields Gaussian profiles of pressure and density for isothermal vertical profiles, and a linear profile for the Brunt-V\"{a}is\"{a}l\"{a} frequency (see BM05):
\begin{mathletters}\label{linear_g}
\begin{eqnarray}
g(z) &=& \Omega_0^2\, z \quad\mathrm{for}~|z|\ll R_0,\\
\bar{P}(z)/(\mathcal{R}T_0) &=& \bar{\rho}(z) = \rho_0\exp(-z^2/2H^2),\quad\quad H\equiv \sqrt{\mathcal{R}T_0}/\Omega_0 = c_s/\Omega_0,\\
N(z) &=& \sqrt{\mathcal{R}/C_P}\, \Omega^2_0|z|/c_s =  \sqrt{(\gamma-1)/\gamma}\, \Omega_0|z|/H, \label{N}
\end{eqnarray}
\end{mathletters}
where $c_s\equiv\sqrt{\mathcal{R}T_0}$ is the isothermal sound speed.\footnote{Historically, the isothermal sound speed has been a convenient velocity scale because $c_s=H\Omega$ for thin Keplerian disks that have uniform temperature in the vertical direction.  But of course, actual sound waves travel at the adiabatic sound speed even if the background is isothermal.} 

As we shall demonstrate, the essential tuning parameter for the ZVI is the Brunt-V\"{a}is\"{a}l\"{a} frequency, so for this paper, rather than a vertical gravity that is linear in $|z|$,  we use a constant vertical gravity, which yields exponential profiles of pressure and density for isothermal vertical profiles, and a constant Brunt-V\"{a}is\"{a}l\"{a} frequency $N_0$:
\begin{mathletters} \label{constant_g}
\begin{eqnarray}
g &=& g_0, \label{g} \\
\bar{P}(z)/(\mathcal{R}T_0) &=& \bar{\rho}(z) = \rho_0\exp(-z/H_0),\quad\quad H_0\equiv \mathcal{R}T_0/g_0 = c_s^2/g_0, \\
N_0 &=& \sqrt{\mathcal{R}/C_P}\, g_0/c_s = \sqrt{[(\gamma-1)/\gamma](g_0/H_0)}.
\end{eqnarray}
\end{mathletters}
In BM05, we previously showed that ZVI is present even when $N(z)$ is spatially varying.  In this paper, by focusing on constant $N_0$, we will be able to probe what thresholds of stable stratification are required to trigger ZVI.  However, this restriction prevents us from exploring how gravity waves and turbulence generated by ZVI will propagate into the nearly unstratified midplane; we defer that important issue to future work.

\subsection{Boundary Conditions and Numerical Resolution} \label{sec:parameters}
All of the simulations in this paper are periodic in the azimuthal or stream-wise direction $y$, and use shearing box boundary conditions in the radial or cross-stream direction $x$ \citep{goldreich65b,marcuspress,rogallo81}.  In the pseudo-spectral calculations, we have performed computations in which the top and bottom boundary conditions are treated by ({\it a}) mapping the boundaries  to $z\rightarrow\pm\infty$ (as in BM05), ({\it b}) setting the vertical velocity to zero (\ie {\it rigid} boundary conditions), or ({\it c}) forcing all variables to be spatially periodic in $z$.  When choice ({\it c}) is used, it is necessary to include a small artificial damping layer adjacent to the top and bottom boundaries to smooth out the inherent non-periodicity of $\bar{\rho}(z)$ and to prevent spurious reflections of inertio-gravity waves and other unphysical effects.  All three of these boundary conditions produce quantitatively similar solutions.  However, option ({\it c}) leads to a triply-periodic and computationally fast code, and thus was used for exploring parameter space for almost all the pseudo-spectral simulations reported in this paper.  The pseudo-spectral domain size was $H_0^3$ and was resolved with $256^3$ spectral modes.  The algorithms used in the pseudo-spectral calculations have no inherent dissipations, so an explicit, weak hyperviscosity is required. 

{\it Athena} is a Godunov finite-volume code that can solve the fully compressible hydrodynamic equations in a shearing box \citep{gs08,stone08,sg10}.   The {\it Athena} simulations used a computational domain of size $H_0^3$ resolved with $256^3$ grid points. The vertical velocity was set to zero at the top and bottom boundaries.  Note that the {\it Athena} code has no explicit dissipation, but its finite-volume algorithms are inherently dissipative.  We found that lower resolution simulations with {\it Athena} (\eg $128^3$ grid points) did not develop ZVI.

\subsection{Choices of Parameters and Dimensionless Numbers} \label{sec:dimensionless}
There are three (reciprocal) timescales of interest: the rotation rate $\Omega_0$, the shear rate $\sigma\equiv -q\Omega_0$, and the Brunt-V\"{a}is\"{a}l\"{a} frequency $N_0$. We consider only Keplerian shear, $q=1.5$ (with the exception of the calculations reported in Fig.~\ref{fig:balbus}). We define the dimensionless ratio $\beta\equiv N_0/\Omega_0$.  For a PPD with linear vertical gravity and no vertical temperature gradient, $\beta(z)=0.63|z|/H_0$.  In BM05, we observed that vortices naturally formed in stratified regions above and below the midplane: $1\lesssim|z|/H_0\lesssim4$, corresponding to $0.63\lesssim\beta\lesssim2.5$.  For constant gravity $g_0$, one can show from Eq.~(\ref{constant_g}) that:
\begin{equation}
c_s = \sqrt{\gamma/(\gamma-1)} \, \beta \, H_0 \, \Omega_0 = \sqrt{5/2} \,\, \beta  \, H_0 \, \Omega_0. \label{home} 
\end{equation}
With constant gravity, the only independent length scale in either the fully compressible or anelastic equations is the gas pressure scale height $H_0$. Therefore, the size of the computational domain $(L_x, L_y, L_z)$  introduces three more dimensionless parameters. In this study, we set $L_x/H_0=L_y/H_0=L_z/H_0 =1$. 

\section{EVIDENCE OF INSTABILITY IN PROTOPLANETARY DISKS WITH VERTICAL GRAVITY} \label{sec:evidence}

Unless otherwise stated, in this section, all times are reported in units of the local orbital period $\tau_{orb}\equiv 2 \pi/\Omega_0$ or local ``years''; all lengths in units of $\Delta\equiv(\beta/3\pi)L_y$ (see Eq.~\ref{delta}); all velocities in units of $\Omega_0H_0$, and all energies per unit mass in terms of $(\Omega_0H_0)^2$.   Table \ref{T:parameters} summarizes the parameters for the figures in this section.  The domain size and resolution for all calculations were $L_x=L_y=L_z=H_0=3\pi/\beta$ with $256^3$ spectral modes, or $256^3$ grid points for {\it Athena} calculations.  All flows were for Keplerian profiles $q=1.5$ (with the exception of Figs.~\ref{fig:balbus} which varied $q$).  The continuity equation was either fully compressible (listed in the bottom row with a C), or was anelastic (listed with an A). Figs.~4, 5, and~7 correspond to the same simulation; Figs. 3b, 6, 10, and 11b correspond to the same simulation; and Figs.~3a (with the solid curve), 8, 11a correspond to the same simulation.  For reference, the Mach number of the flows in this section are computed with the isothermal sound speed (rather than the adiabatic sound speed): $Ma\approx 0.63 \, \beta^{-1}$ multiplied by the dimensionless velocity.
 
 \begin{table}[ht]
\begin{threeparttable}
\begin{center}
\begin{tabular}{|l|c|c|c|c|c|c|}
\hline
{\bf Figures} & {\bf  2a}\tnote{$\dagger$} & {\bf 2b} & {\bf 3a, 8, 11a} & {\bf 3b, 6, 10, 11b} & {\bf 4, 5, 7} & {\bf 9} \\
\hline
\hline
{\bf $\beta\equiv N_0/\Omega_0$} & {0} & {0} & {2} & {2} & {1}  & {2} \\
\hline
{\bf Initial rms} & & & & & &   \\
{\bf velocity} & 0.1& 0.02& 0.01& 0.02& 0.01&  0.01\\
\hline
{\bf Equation} & & & & & &  \\
{\bf of state} & C & A & A & C & A &  A  \\
\hline
\end{tabular}
\begin{tablenotes}
\item [$\dagger$] Fig.\ 2a is a copy of Fig.\ 1 from \citet{balbus96}.
\end{tablenotes}
\caption{\label{T:parameters} Parameters Values. The domain size and resolution for all calculations were $L_x=L_y=L_z=H_0=3\pi/\beta$ with $256^3$ spectral modes or grid points.  All flows were for Keplerian profiles $q=1.5$ (with the exception of Fig.~\ref{fig:balbus} which varied $q$).  The continuity equation was either fully compressible (C) or anelastic (A).  Figs.~4, 5, and~7 are for the same simulation; Figs. 3b, 6, 10, and 11b are for the same simulation; and Figs.~3a (with solid curve), 8, and 11a are for the same simulation. 
}
\end{center}
\end{threeparttable}
\end{table}
 
\subsection{Temporal Growth and Decay of Initial Energy Fluctuations} \label{sec:evidence.1}

One of the most cited pieces of evidence that PPDs are stable to purely-hydrodynamic instabilities is given by Fig.~1 in \citet{balbus96}, which we copy below as Fig.~\ref{fig:balbus}a. The figure shows the growth or decay of kinetic energy fluctuations as a function of time for various values of $q$, where $q$ is defined in Eq.~(\ref{q}) and where the  kinetic energy fluctuation is defined as $|{\bf v}(x,y,z,t) - \bar{v}_y \, \hat{\bf y}|^2/2$, where $\bar{v}_y \, \hat{\bf y}$ is the steady equilibrium flow in Eq.~(\ref{hydrostatic2}).  The initial-value calculations that produced Fig.~\ref{fig:balbus}a used the fully compressible equations~(\ref{BHS35})--(\ref{BHS35d}),
and were initialized with the steady equilibrium velocity in Eq.~(\ref{hydrostatic2}).  However, this flow was computed with {\it no} vertical gravity and was therefore initialized with the equilibrium that had $\bar{P}=P_0$, $\bar{\rho}=\rho_0$, and $\bar{T}=T_0$. The initial flow was perturbed with small-amplitude noise (with root-mean-square or rms velocity of $0.1$) that had a Gaussian-like energy spectrum as a function of wavenumber.  The equations were solved with periodic boundary conditions in $y$ and $z$, and with shearing box boundaries in $x$.  Because the energy fluctuations all increase in time for $q \ge 2$ and decrease in time for $q < 2$,  Fig.~\ref{fig:balbus}a is used to support  the hypothesis that these flows of ideal gases are stable (unstable) to all purely-hydrodynamic  perturbations when the angular momentum per unit mass of the flow, $R^2 \, \Omega(R) \propto R^{(2-q)}$, increases (decreases) in the radially outward direction. This hypothesis is consistent with Rayleigh's centrifugal stability theorem; however, it must be noted that Rayleigh's theorem was proved only for the case of constant density fluids and not for stratified ideal gases \citep{R17,synge1933}, and therefore, it may not be applicable to astrophysical flows in disks. In particular, initial conditions in which the density is constant prohibit baroclinic instabilities such as ZVI (MPJH13). The curve labeled with ``shr'' in Fig.~\ref{fig:balbus}a corresponds to the case of pure Cartesian shear with the Coriolis and tidal acceleration terms dropped from Eq.~(\ref{BHS35b}), which would be appropriate for fully compressible flow in a channel with cross-stream shear, but no rotation. 

In Fig.~\ref{fig:balbus}b, we reproduce the results of Fig.~\ref{fig:balbus}a with our pseudo-spectral numerical code \citep{barranco06} using the anelastic equations~(\ref{E:anelastic_eqns}) with no vertical gravity.  The initial noise that we added to the equilibrium velocity in all of our numerical experiments was random, homogeneous, and isotropic with a Kolmogorov spectrum so that the kinetic energy per unit mass was $E(k) \propto k^{-5/3}$, where $k$ is the wavenumber.   Our primary reason for reproducing this figure from \citet{balbus96} is to show that we get the same stability results for no vertical gravity with a pseudo-spectral code (vs. finite-volume code) which solve the anelastic equations (vs. the fully compressible equations).  

\unitlength1mm{
\begin{figure}[H]
\begin{picture}(83,86)
\put(-2,2){\epsfig{figure=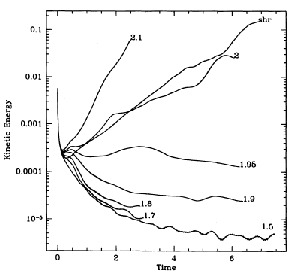,width=88mm}}
\put(90,7){\epsfig{figure=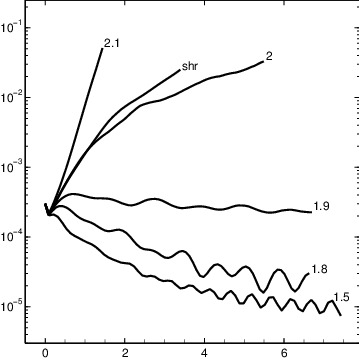,width=75mm}} 
\end{picture}
\put(40,5) {Time}
\renewcommand{\baselinestretch}{1.0}
\caption{\label{fig:balbus} (a) Left panel: Fig.~1 from  \citet{balbus96}, which shows the temporal evolution of the fluctuation kinetic energy per unit mass, for different values of $q$ as defined by Eq.~(\ref{q}). These are fully compressible simulations with $g=0$, $N=0$, $\gamma=5/3$.  The calculation was done with the finite-difference {\it Zeus} code with spatial grid of $63^3$ ($127^3$ for $q=1.5,2$) points. The equilibrium flow had uniform pressure, density, and temperature.  The initial rms velocity amplitude of the noise was $\sim0.1$.    (b) Right panel: Same as left panel, but computed with our pseudo-spectral code and the anelastic equations, rather than the fully compressible equations. The initial noise had an rms velocity of $\sim0.02$.}
\renewcommand{\baselinestretch}{2.0}
\end{figure}}

How does vertical stratification affect stability? The decaying dashed curve in Fig.~\ref{fig:growenergy}a is the time evolution of the kinetic energy fluctuations for a unstratified flow similar to the one as given by the curve in Fig.~\ref{fig:balbus}b with $q=1.5$ and $g=N=\beta=0$, but plotted for a longer time and a smaller value of initial noise.  The solid curve in Fig.~\ref{fig:growenergy}a shows the evolution of kinetic energy fluctuations when vertical gravity is included; here,  $g=g_0 \ne 0$ is constant, $N = N_0 \ne 0$, and $\beta = 2$.  The growth of the kinetic energy fluctuations shows that the flow is unstable with $q=1.5$ when vertical gravity is present.  Fig.~\ref{fig:growenergy}b  shows the kinetic energy growth for the same flow as shown with the solid curve as in
Fig.~\ref{fig:growenergy}a, but computed with the fully compressible equations with {\it Athena}.  The rapid growth of the kinetic energy fluctuations shows that ZVI is not an artifact of the anelastic approximation or spectral codes.

\unitlength1mm{
\begin{figure}[H]
\begin{picture}(166,72)
\put(-2,6){\epsfig{figure=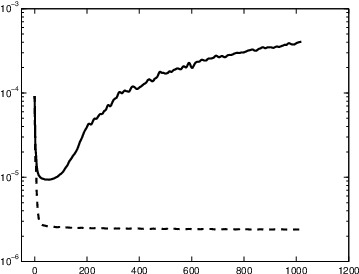,width=86mm}}
\put(-7,27) {\rotatebox{90}{Kinetic Energy}}
\put(39,1) {Time}
\put(88, 6){\epsfig{figure=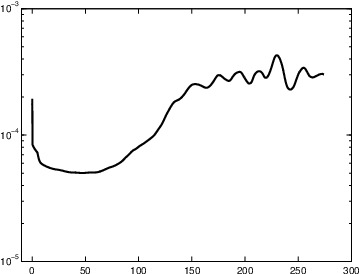,width=86mm}} 
\put(130,1) {Time}
\end{picture}
\renewcommand{\baselinestretch}{1.0}
\caption{\label{fig:growenergy} (a) Left panel: Time evolution of the fluctuation kinetic energy per unit mass for anelastic flows for $q = 1.5$. {\it Dashed curve} -- unstratified initial flow with $g=N=\beta=0$, which is similar to the calculation shown in Fig.~\ref{fig:balbus}b labeled with ``1.5'', but integrated for longer time.  {\it Solid curve} -- The same anelastic flow as shown with the dashed curve but stratified with $g_0 \ne 0$ and $\beta =2$.  Note the difference stratification makes to stability.  The energies have been time-averaged with a window size of 10~yrs.  The time evolution of kinetic energy can be divided into three phases. From $t=0$ to 50~yrs the flow adjusts from the initial condition with most of its initial vorticity dissipated by hyperviscosity. From $t=50$ to $250$~yrs the non-Keplerian kinetic energy fluctuations increase nearly exponentially; during this time, the critical layers are strongly excited (see $\S$\ref{sec:cartoon}), turn into vortex layers, and roll-up into zombie vortices.   From $t=250$~yrs onward, the growth of the non-Keplerian kinetic energy fluctuations slows until it reaches a statistically steady state.  (b) Right panel: Similar initial stratified flow with $\beta=2$ as in the solid curve in the panel on the left, but computed with the fully compressible equations using the {\it Athena} code, and with an rms velocity of the initial noise that is $\sim2$ times larger than the unstable flow on the left.  The growth of the non-Keplerian energy shows that ZVI occurs in fully compressible flows computed with finite-volume codes and is neither an artifact of the anelastic approximation nor of spectral codes. Due to the inherent dissipation in the {\it Athena} code, the growth rate and the late-time non-Keplerian kinetic energy are smaller in panel (b) than in (a).
}
\renewcommand{\baselinestretch}{2.0}
\end{figure}
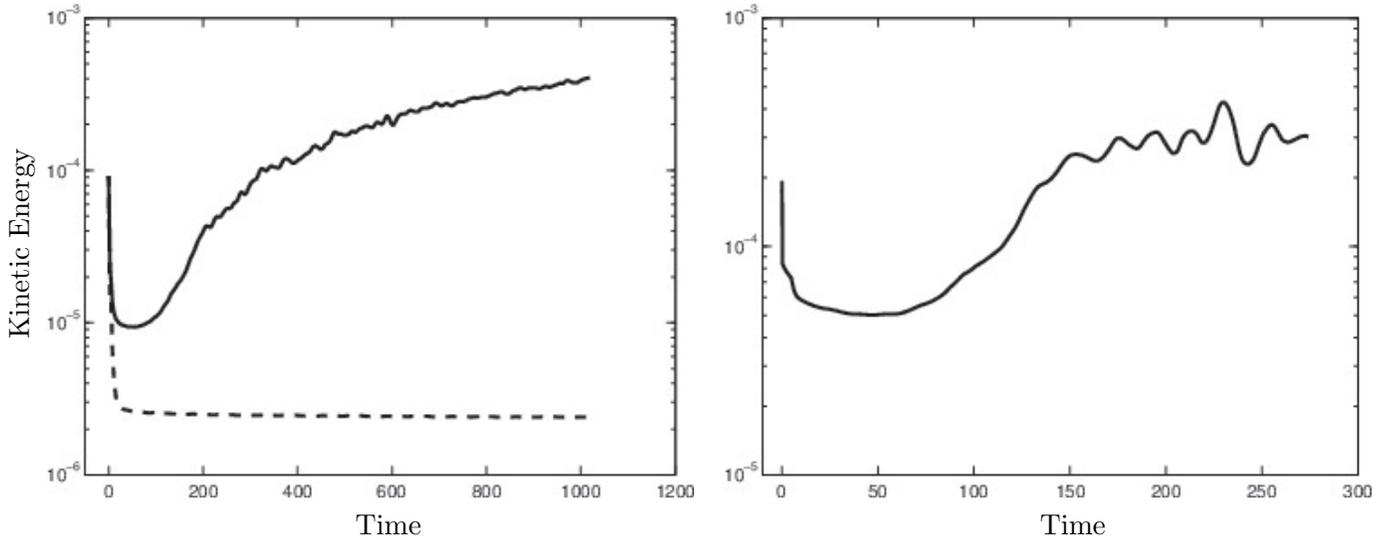}

\unitlength1mm{
\begin{figure}[H]
\begin{center}
\begin{picture}(0,160)
\put(-90,92){\epsfig{figure=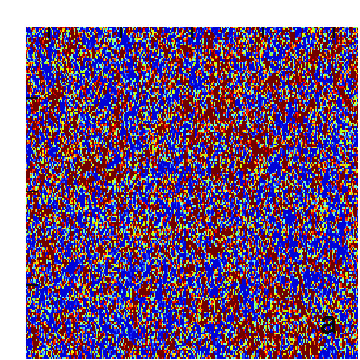,width=90mm}}   
\put(-90,  1){\epsfig{figure=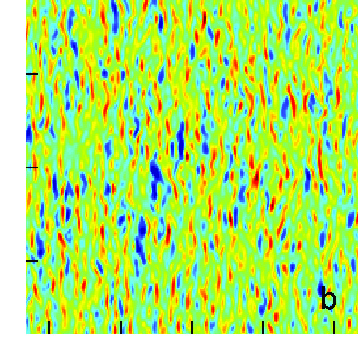,width=90mm}}  
\put(1, 92){\epsfig{figure=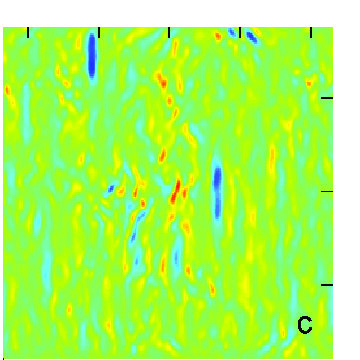,width=90mm}} 
\put(1,  1){\epsfig{figure=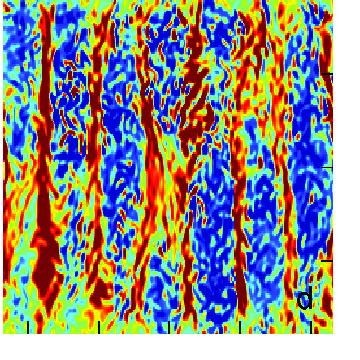,width=90mm}} 
\put(-87,71) {$2$}
\put(-87,47.5) {$0$}
\put(-90,24.5) {$-2$}
\put(-87,156) {$2$}
\put(-87,132.5) {$0$}
\put(-90,109.5) {$-2$}
\put(-89.5,91) {$z$}
\put(-81.5,3.5) {$-4$}
\put(-63.5,3.5) {$-2$}
\put(-43.0,3.5) {$0$}
\put(-25,3.5) {$2$}
\put(-7.5,3.5) {$4$}

\put(4.5,3.5) {$-4$}
\put(22,3.5) {$-2$}
\put(42.5,3.5) {$0$}
\put(60.5,3.5) {$2$}
\put(78,3.5) {$4$}
\put(0,1) {$x$}
\end{picture}
\renewcommand{\baselinestretch}{1.0}
\caption{ \label{fig:XZ} Evolution of  $Ro(x,y,z,t)$ in the $x-z$  plane. 
The figure is cropped so as to not show the damping regions at the vertical boundaries. The anelastic flow has $\beta=1$ and is perturbed initially with homogeneous, isotropic noise with a Kolmogorov spectrum.  The color-map ranges from $-0.25$ to $0.25$, with blue [red] for anticyclones [cyclones] with  $Ro<0$ [$Ro > 0$].  Green corresponds to $Ro=0$. 
(a) $t=0$~yrs: Relative vorticity is dominated by the smallest length scales, so the image is pixelated at the resolution length. 
The color scale is over-saturated in panel~(a).  The actual extremum value of the Rossby number for the initial condition is $\max(|Ro|)=2.4$.
(b) $t=2.5$~yrs: The Keplerian background shear has stretched the relative vorticity and elongated it in the $y$ direction; the flow's energy 
has cascaded to small scales where it is dissipated by hyperviscosity causing the amplitude of $|Ro(x,y,z,t)|$ to decrease by an order of magnitude from its initial value.
(c) $t=50.9$~yrs: Mergers of small anticyclones in a reverse cascade of energy to larger scales has led to a few surviving, but spatially scattered, anticyclones.
(d) $t=1370$~yrs: Zombie turbulence and zombie vortices with $ Ro \simeq-0.3$ fill the domain. The near spatial periodicity of the flow in $x$, with a wavenumber (in this case $\sim$$7$) slightly smaller than $L_x/\Delta$ is one of the signatures that makes zombie turbulence easy to identify.  
The online version has a link to an MPEG movie corresponding to this figure.
}
\renewcommand{\baselinestretch}{2.0}
\end{center}
\end{figure}
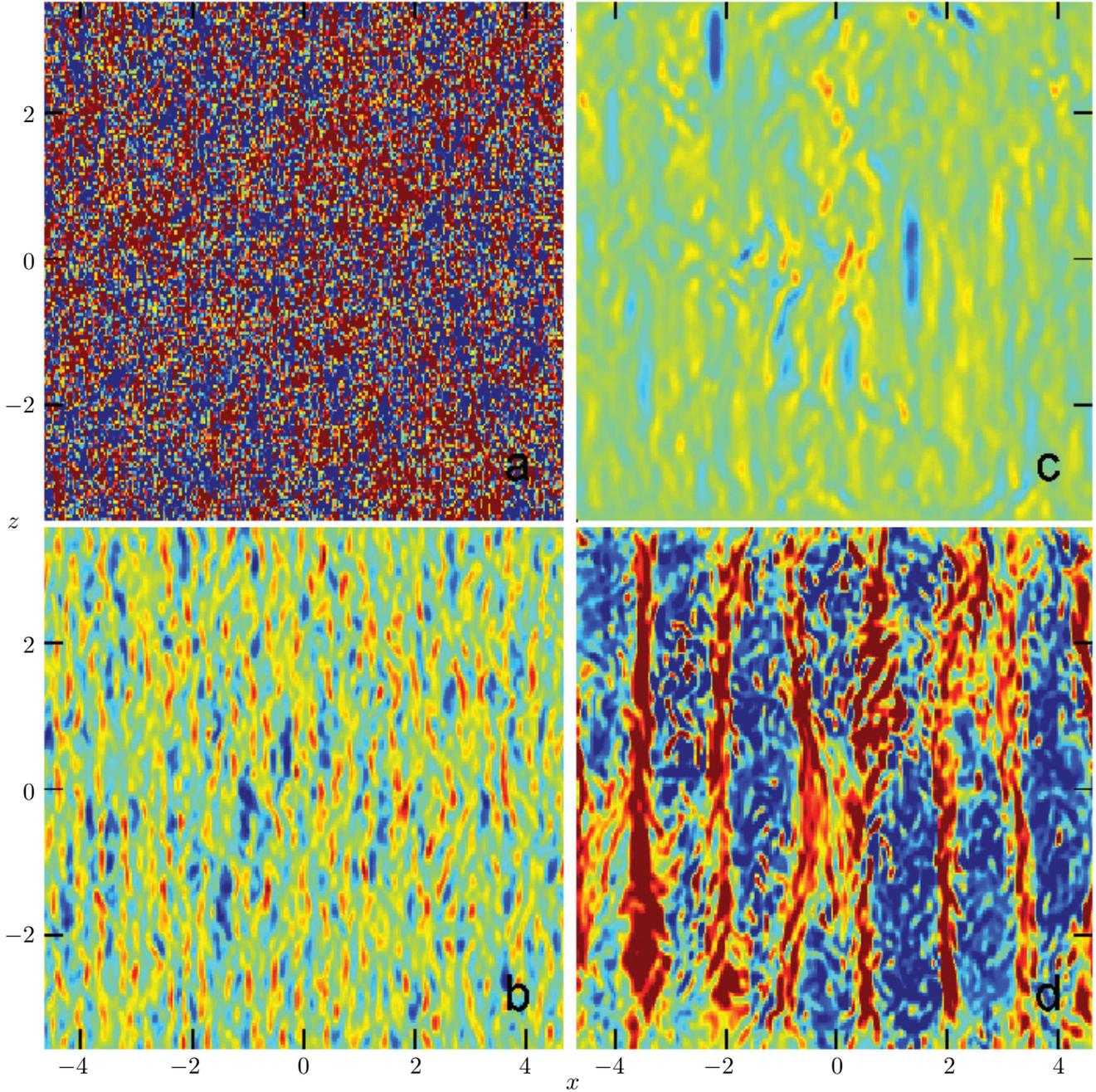}

\unitlength1mm{
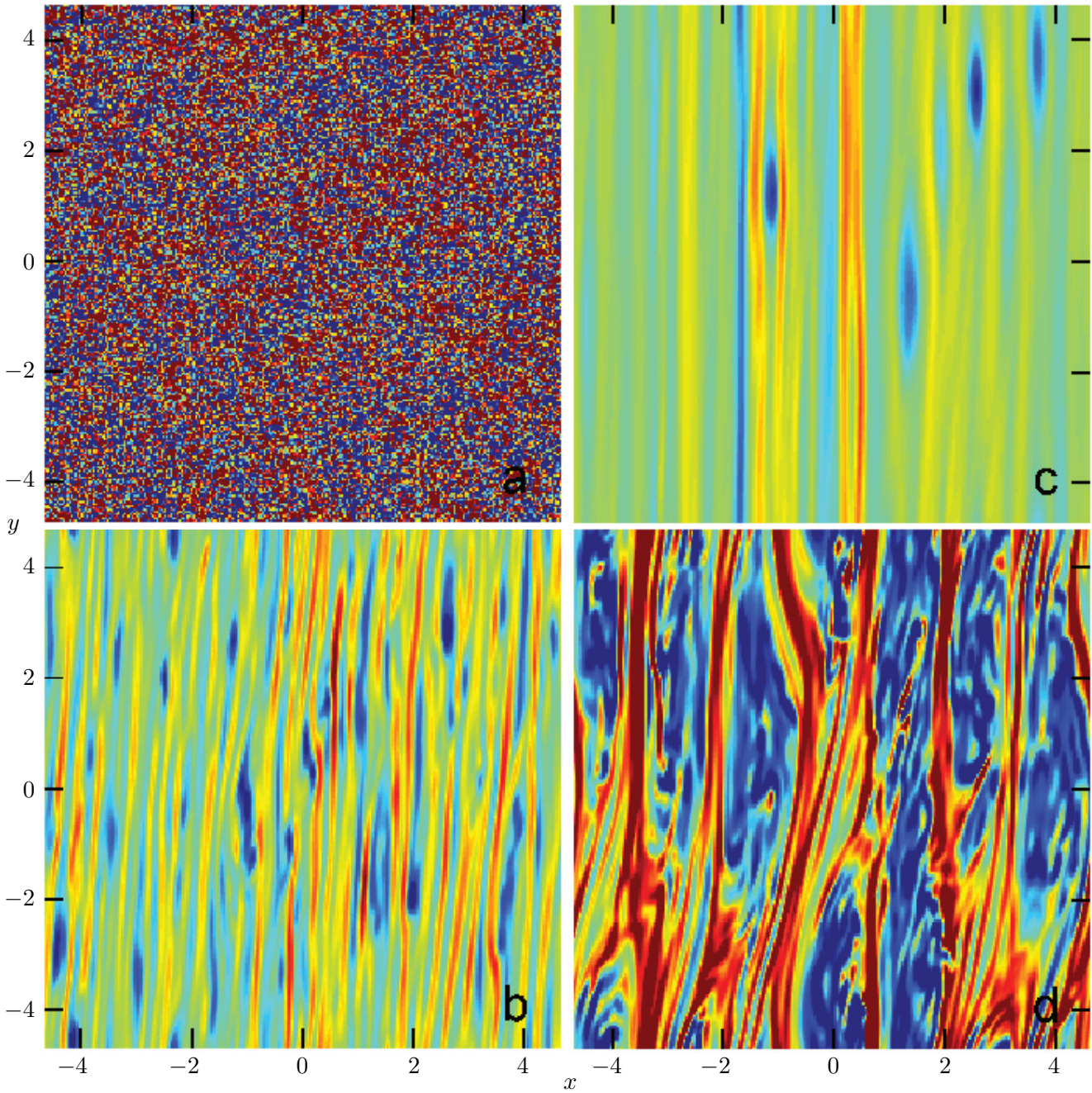
\begin{figure}[H]
\begin{center}
\begin{picture}(0,176)
\put(-90,92){\epsfig{figure=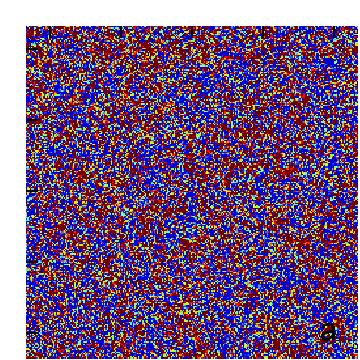,width=90mm}}   
\put(-90,  1){\epsfig{figure=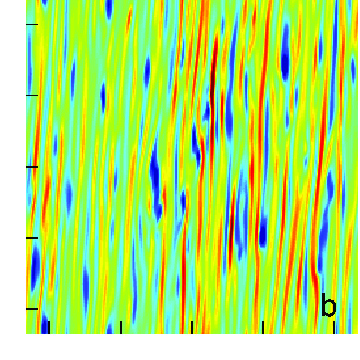,width=90mm}}   
\put(1, 91.7){\epsfig{figure=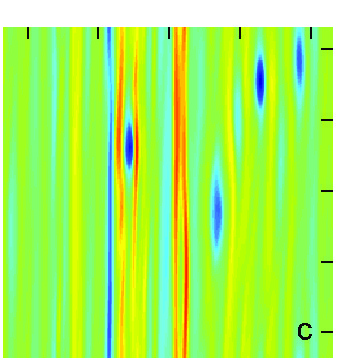,width=90.5mm}} 
\put(1,  0.7){\epsfig{figure=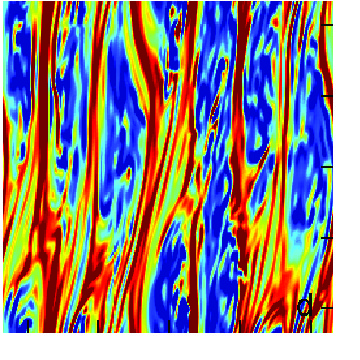,width=90.5mm}}  
\put(-87,84) {$4$}
\put(-87,66) {$2$}
\put(-87,48) {$0$}
\put(-90,30.5) {$-2$}
\put(-90,12.5) {$-4$}
\put(-87,169) {$4$}
\put(-87,151) {$2$}
\put(-87,133) {$0$}
\put(-90,115.5) {$-2$}
\put(-90,98) {$-4$}
\put(-89.5,91) {$y$}
\put(-81.5,3.5) {$-4$}
\put(-63.5,3.5) {$-2$}
\put(-43.0,3.5) {$0$}
\put(-25,3.5) {$2$}
\put(-7.5,3.5) {$4$}
\put(4.5,3.5) {$-4$}
\put(22,3.5) {$-2$}
\put(42.5,3.5) {$0$}
\put(60.5,3.5) {$2$}
\put(78,3.5) {$4$}
\put(0,1) {$x$}
\end{picture}
\renewcommand{\baselinestretch}{1.0}
\caption{ \label{fig:XY} 
Same as Fig.~\ref{fig:XZ} but in the $x$-$y$ plane at $z=0$.
Panel (a)  looks like  Fig.~\ref{fig:XZ}a because the initial noise is isotropic and homogeneous.
The online version has a link to an MPEG movie corresponding to this figure.
}  
\renewcommand{\baselinestretch}{2.0}
\end{center}
\end{figure}}

\unitlength1mm{
\begin{figure}[H]
\begin{center}
\begin{picture}(0,92)
\put(-92,1){\epsfig{figure=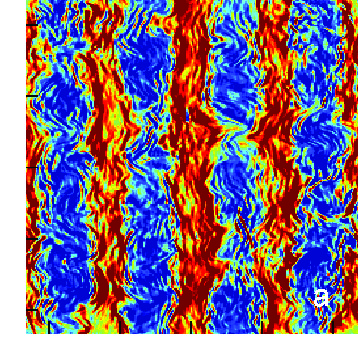,width=90mm}}  
\put(1,  1){\epsfig{figure=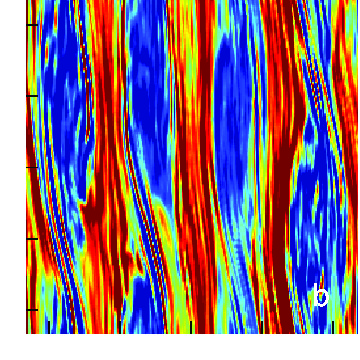,width=90mm}}  
\put(-89,84) {$2$}
\put(-89,66) {$1$}
\put(-89,47.5) {$0$}
\put(-92,30) {$-1$}
\put(-92,12.5) {$-2$}
\put(-93,48) {$z$}
\put(-83.5,3.5) {$-2$}
\put(-65.5,3.5) {$-1$}
\put(-45.0,3.5) {$0$}
\put(-27,3.5) {$1$}
\put(-9.5,3.5) {$2$}
\put(-45,0) {$x$}
\put(4,84) {$2$}
\put(4,66) {$1$}
\put(4,47.5) {$0$}
\put(1,30) {$-1$}
\put(1,12.5) {$-2$}
\put(0,48) {$y$}

\put(10.0,3.5) {$-2$}
\put(28,3.5) {$-1$}
\put(48,3.5) {$0$}
\put(66,3.5) {$1$}
\put(84,3.5) {$2$}
\put(48,0) {$x$}
\end{picture}
\renewcommand{\baselinestretch}{1.0}

\caption{\label{fig:athena}
Point-wise Rossby number $Ro(x, y, z, t)$ for the fully compressible $\beta=2$ flow shown in Fig.~\ref{fig:growenergy}b at $t =274$~yrs using the same colormap range as in Figs.~\ref{fig:XZ} and~\ref{fig:XY}. No damping at the vertical boundaries is used in this simulation so the flow is uncropped and shows the full computational domain.  
(a) As in Fig.~\ref{fig:XZ}d.
(b) As in Fig.~\ref{fig:XY}d. It is possible that at this time, the flow has not reached a statistically steady state, and the pixelated structures in panel (b) indicate that the numerical dissipation in {\it Athena} is significant on small scales. However, the point we emphasize here is that zombie turbulence occurs in fully compressible flows computed with a finite-volume code, and the turbulence looks qualitatively similar to that in anelastic flows computed with a spectral code.  The online version has links to MPEG movies corresponding to this figure.
}
\renewcommand{\baselinestretch}{2.0}
\end{center}
\end{figure}
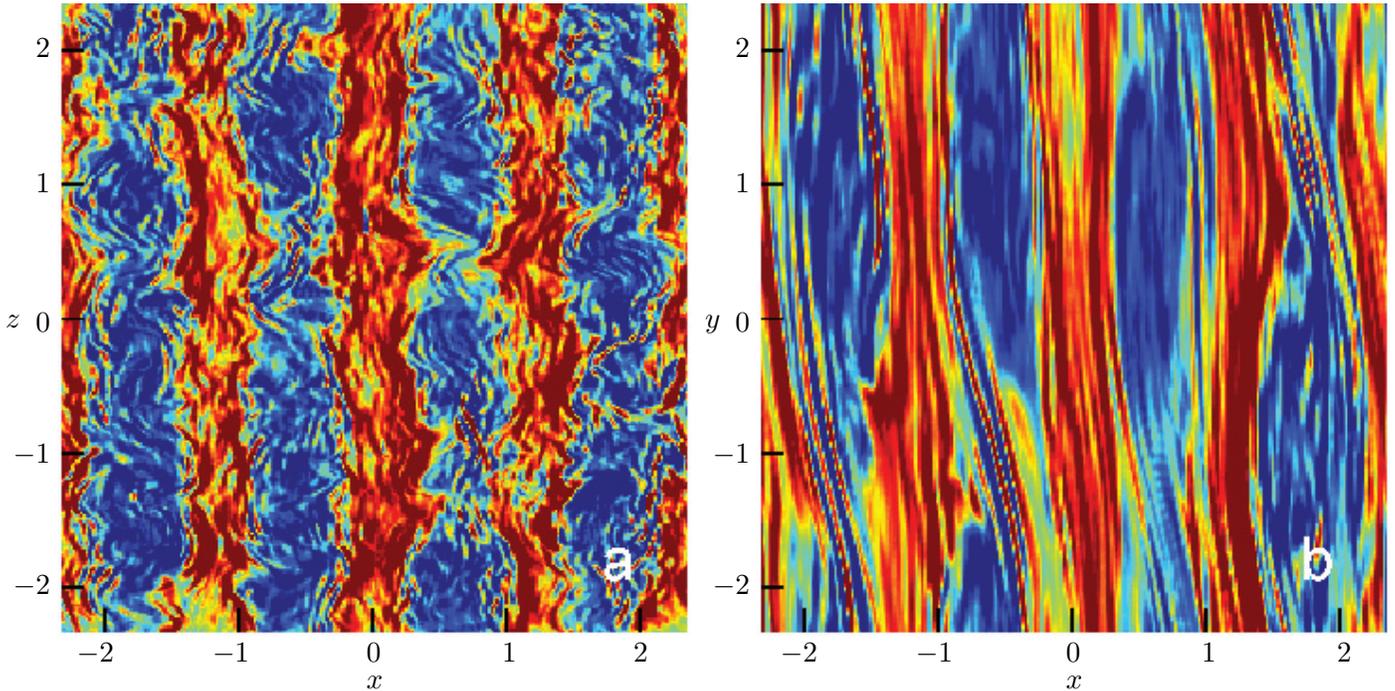}

\subsection{The Vertical Vorticity of the Zombie Vortex Instability} \label{subsec:reverse}

The growth of kinetic energy fluctuations in Figs.~\ref{fig:growenergy} is evidence of instability, but spatial plots of the relative vorticity, defined as $\bfome \equiv \bfDel \times ({\bf v} - \bar{\bf v}) = \bfDel \times {\bf v}  + q \, \Omega_0 \, \hat{\bf z}$, are more useful in illustrating ZVI and its nonlinear evolution. The point-wise dimensionless Rossby number, defined as the ratio of relative vorticity to the global Keplerian vorticity $Ro(x,y,z,t) \equiv [\hat{\bf z} \cdot \bfome (x, y, z, t)]/(2 \Omega_0)$, is a better diagnostic of late-time zombie turbulence than, say, the Mach number, because the amplitude of the Mach number at late-times is sensitive to the values of $L_y/L_x$ and $H_0/L_x$ (which are chosen arbitrarily in our simulations), whereas the Rossby number is not -- see \S\ref{subsec:mach}.  The simulations in this section are initialized by random perturbations which are closer to the expected conditions in a PPD rather than by an {\it ad hoc} isolated initial vortex as was done in the simulations in BM05 and MPJH13. Fig.~\ref{fig:XZ} shows $Ro(x,y,z,t)$ for an anelastic flow in the $x-z$ plane at four different times and at an arbitrary stream-wise location in $y$. (Because the equations, boundary conditions and initial conditions are invariant under translation in $y$, the flow in all  $x-z$ planes is statistically the same for all time.)   Fig.~\ref{fig:XY} shows $Ro(x,y,z,t)$ for the same flow in the $x-y$ plane at $z=0$, which is  midway between the upper and lower boundaries. The parameter values, initial conditions and boundary conditions of the flow in Figs.~\ref{fig:XZ} and \ref{fig:XY} are similar to the flow shown by the solid curve in Fig.~\ref{fig:growenergy}a with the difference that $\beta= 1$, rather than 2.  The initial perturbing noise has a power-law spectrum with random velocity phases, so there are no inherent length scales or coherent features of any kind in the initial perturbation. For Kolmogorov noise, the energy is dominated by Fourier modes with the smallest spatial wave numbers, while the relative vorticity field is dominated by Fourier modes with the largest spatial wavenumbers. Thus, Figs.~\ref{fig:XZ}a and~\ref{fig:XY}a are dominated by the smallest length scales (which are equal to the spatial resolution of the calculation, $L_x/256$ in each direction).  Much of the initial vorticity is quickly stretched and elongated in the stream-wise direction by the Keplerian shear, as shown in Figs.~\ref{fig:XZ}b and~\ref{fig:XY}b. The stretching cascades the initial kinetic energy to spatial wavenumbers with higher Fourier modes where it is quickly dissipated by hyperviscosity (similar to Fig.~\ref{fig:growenergy}a at $t \simeq 50$).  In competition with the stretching, there is a weaker reverse cascade of energy from small to large length scales due to the fact the fluid is rapidly rotating. Here, the reverse cascade manifests itself via vortex mergers. In a rapidly rotating flow without a background shear, like-signed vortices of both signs merge, but when a strong background shear is also present, vortices with the same sign as  the background shear merge into larger vortices with shapes that are elongated in the stream-wise $y$ direction, while vortices  with a sign opposite to the background shear flow are stretched into thin stream-wise-oriented filaments and layers \citep{marcus93}.  The shear in a PPD is  anticyclonic, and therefore anticyclones merge, while cyclones are stretched.\footnote{Relative vorticity is defined as cyclonic [anticyclonic]   when $\omega_z$ and $\Omega_0$ have the same [opposite] signs, or equivalently, when $Ro(x,y,z,t) > 0$  [$Ro(x,y,z,t) < 0$].}  The result of the competition between the forward energy cascade (resulting in energy dissipation) and the reverse energy cascade (resulting in the mergers of anticyclones)  is visible in Figs.~\ref{fig:XZ}c and~\ref{fig:XY}c at  time $t=51$~yrs; the flow is dominated by a few scattered anticyclones.  Incidentally, this flow looks very much like the initial condition of a single isolated anticyclone that we studied in MPJH13 and illustrated in Fig.~\ref{F:zombie_cartoon}. 

By $51$~yrs, ZVI is well underway in Figs.~\ref{fig:XZ} and~\ref{fig:XY}.  The scattered anticyclones which emerged from the reverse cascade have now excited neighboring critical layers to form dipolar vortex layers.  The critical layers themselves are easy to identify in plots of $v_z(x,y,z,t)$ (not shown), but difficult to observe in plots of $Ro(x,y,z,t)$. However, the vortex layers that the critical layers create are easily identifiable in plots of $Ro(x,y,z,t)$ because each dipolar sheet appears as a pair of opposite-signed ``stripes'' in the $x$-$y$ plane (with the critical layer sandwiched between them -- Fig.~\ref{F:zombie_cartoon}c).  In an initial-value simulation with an initial perturbation consisting of noise, it is sometimes difficult to see the  dipolar vortex layers  due to the fast instability of the anticyclonic sheet. However in carefully controlled simulations consisting of a single perturbing vortex, rather than noise, dipolar layers can be found easily, as shown in Fig.~1a in MPJH13.  As discussed in $\S$\ref{sec:cartoon}, the layers of cyclonic vorticity aligned in the stream-wise direction are linearly stable. In contrast, layers of anticyclonic vorticity aligned in the stream-wise direction are linearly unstable and roll-up into stable anticyclones. 

At late times (Figs.~\ref{fig:XZ}d and~\ref{fig:XY}d), the flow has reached a statistically steady state of zombie turbulence.  One can see by eye, especially in Fig.~\ref{fig:XY}d, that the flow has formed a pattern with cross-stream or $x$ wavenumber of $\sim$$7$.  The relative vorticity, although very  turbulent, has developed  spatial coherence and is neither homogeneous nor isotropic. The cyclonic vorticity has formed approximately 2-dimensional layers  that are approximately aligned in the $y$-$z$ planes.  Between these planes are approximately ellipsoidal turbulent anticyclones. At intermediate  times in our calculation, we would  expect a  dominant $x$-wavenumber of $L_x/\Delta = 3 \pi \simeq 10$, and at times between that of Figs.~\ref{fig:XY}c and that of Fig.~\ref{fig:XY}d,  the dominant $x$-wavenumber is indeed $\sim$$10$.  However, at later times, the vortices grow in size, run into each other in the $x$ direction, and  become larger.  Due to  mergers, the late-time $x$-diameters of the vortices and the average spacing in $x$ between the cyclonic sheers both become slightly larger than $\Delta$. Therefore, the dominant  wavenumber in $x$ becomes slightly less than $L_x/\Delta$.  We have carried out initial-value calculations of zombie turbulence  in anelastic flows and Boussinesq flows (MPJH13) for a wide variety of parameters, and the $Ro(x,y,z,t)$ always look like the pattern in Figs.~\ref{fig:XZ}d and~\ref{fig:XY}d. Our one fully compressible calculation
(Fig.~\ref{fig:athena}) also shows this same pattern at late times. In Fig.~\ref{fig:athena}, $\beta =2$ and $L_y=L_x$, so we would expect  that $L_x/\Delta \simeq 5$, which argues that the $x$ wavenumber that initially dominates the flow is $5$. This wavenumber agrees with this simulation at intermediate times.  At late times, the wavenumber in Fig.~\ref{fig:athena} has been reduced to $\sim$$4$ by vortex mergers and reverse energy cascades. Note that the two flows shown in Figs.~\ref{fig:exam2} and~\ref{fig:exam3}, also have $\beta =2$ and $L_y=L_x$, and although we have not included figures of their  $Ro(x,y,z,t)$, these flows also have dominant $x$-wavenumbers of $\sim$$5$ at intermediate times and $\sim$$4$ at late times.  Thus, the main signature of zombie turbulence that differentiates it from other types of turbulence in shear flows that are triggered by finite-amplitude instabilities is that the turbulence in ZVI is never homogeneous or isotropic; it is dominated by cyclonic layers and elongated anticyclones aligned in the stream-wise direction of the background shear, and the pattern of layers and anticyclones has a dominant $x$ wavenumber slightly less than $L_x/\Delta$. 

The aspect ratio $\chi$ (defined as the stream-wise diameter of a vortex in the $y$ direction divided by its cross-stream diameter in $x$) of the zombie vortices is approximately the same as the laminar vortices studied by \citet{mooresaffman} and investigated in BM05 in the PPD context, where $\chi$ is given implicitly by  
\begin{equation}
-\frac{\omega_z}{q \, \Omega_0} \equiv \frac{-2 Ro}{q} = \frac{(\chi+1)}{(\chi-1)}\frac{1}{\chi}. \label{saffman}
\end{equation} 
The Moore-Saffman relation was derived for  a steady two-dimensional vortex with uniform relative vorticity embedded in flow with uniform shear.  However, the turbulent zombie vortices have aspect ratios similar to that in Eq.~(\ref{saffman}) because the relation is the quantification of the fact that a large relative vorticity tends to make a vortex ``round'' and a large background shear tends to elongate a vortex is its stream-wise direction.  At late times the characteristic magnitude of $Ro$ of the anticyclones in zombie turbulence is always $\sim$-0.25, so  regardless of the parameters of the flow, Eq.~(\ref{saffman}) shows that in a Keplerian disk, the aspect ratios $\chi$  of zombie vortices are between 4 and 5.

\subsection{Space-Filling Properties of Zombie Turbulence} \label{subsec:mach}

Figures~\ref{fig:exam1} --~\ref{fig:exam4} show how the root-mean-square (rms) Rossby numbers $Ro_{rms}(t)$ and rms Mach number $M\!a_{rms}(t)$ evolve in time for three anelastic flows and one fully compressible flow.  The flows in Figs.~\ref{fig:exam2} and~\ref{fig:exam3} have the same flow parameters, with the only difference being the rms value of their initial noise.  All of the statistical properties, including $Ro_{rms}(t)$ and $M\!a_{rms}(t)$, of the late-time flows in Fig.~\ref{fig:exam2} and~\ref{fig:exam3} are nearly the same, which suggests that they evolve to a common attracting solution that is independent of the initial conditions. For all four flows in Figs.~\ref{fig:exam1} --~\ref{fig:exam4}, the values of  $Ro_{rms}(t)$ initially plummet due to small-scale hyperviscosity (or numerical diffusion in {\it Athena}), but then grow after  ZVI develops, and eventually reach a plateau with $Ro_{rms}(t)$ between $0.2$ and $0.3$.

The $Ro_{rms}(t)$ is the ratio of the strength of the vorticity in the turbulence to that of the unperturbed Keplerian flow, so $|Ro_{rms}(t)|\sim0.2 -0.3$ indicates large amplitude turbulence.  We have no rigorous explanation for why the Rossby numbers evolve to these values in {\it all} of our simulations that produce zombie turbulence, regardless of parameter values. However, this range of values of $|Ro_{rms}|$ is consistent with the picture that we presented in MPJH13 and Fig.~\ref{F:zombie_cartoon} of how ZVI spreads throughout the computational domain after it starts at one or more discrete locations.  We showed that after the first critical layer is excited, it creates a vortex dipolar layer that rolls up into the first zombie vortex. Then, that first zombie vortex triggers an  instability in an adjacent critical layer in an unperturbed region of the flow,  which then launches a new vortex layer and zombie vortex that triggers the instability in its adjacent critical layer, and so on {\it ad infinitum}. However, this scenario requires that the vortices that make up the zombie turbulence have sufficiently large amplitudes to trigger the finite-amplitude ZVI in unperturbed regions of the flow.

It is more common among astrophysicists to measure the amplitude of turbulence with the Mach number, rather than the Rossby number, so it is somewhat disconcerting that the late-time values of  $M\!a_{rms}(t)$ in our calculations depend on the values of $H_0/L_y$, and  that  the rms Mach numbers can be made big or small by adjusting the size of the computational box. It is also discouraging (to those who might believe that zombie turbulence can transport angular momentum in PPDs) that  $M\!a_{rms}(t)$ is only $\sim$~$10^{-2}$ in our calculations. We now show that the dependence of  $M\!a_{rms}$ on the size of the computational box is easily explained, and that in a PPD we expect  $M\!a_{rms}$ to be equal to $|Ro_{rms}|$  at late times.  Note that the rms Rossby number is approximately
\begin{equation}
Ro_{rms} \simeq V_{eddy}(L_{\omega})/(L_{\omega} \, \Omega_0), \label{rmsro}
\end{equation}
where $L_{\omega}$ is the characteristic length scale of the flow where the vorticity has its maximum value and $V_{eddy}(L)$ is the characteristic velocity of a turbulent eddy of diameter $L$.\footnote{In the literature of turbulence, an ``eddy'' is not the same as a Fourier mode, rather it is a component of the velocity made from a {\it band} of Fourier modes. The kinetic energy per unit mass of an eddy is defined in terms of the differential kinetic energy spectrum $E(k)$ of the flow, where $k$ is the Fourier wavenumber. The kinetic energy per unit mass of the total flow is $\int^{\infty}_{0} \,\, E(k) \,\, dk$, and the kinetic energy per unit mass of an eddy of diameter $L$ is defined as $\int^{2 \pi/L}_{\pi /L} \,\, E(k) \,\, dk \equiv [V_{eddy}(L)]^2/2$. The last equivalence is what is used to defined the eddy velocity $V_{eddy}(L)$. Note that an eddy of diameter $L$ is made up of all velocity Fourier modes with wavenumbers $2 \pi/(2L) < k <  2\pi/L$. Kolmogorov turbulence has an energy spectrum $E(k) \propto k^{-5/3}$, so in Kolmogorov turbulence, $V_{eddy}(L)$ is proportional to $L^{1/3}$ -- see \citet{tennekes1972first}.}
The rms Mach number is approximately
\begin{equation} 
M\!a_{rms} \simeq V_{eddy}(L_v)/c_s, \label{rmsma}
\end{equation} 
where $L_v$ is the characteristic length scale of the flow where the velocity has its maximum value.  In a Kolmogorov energy spectrum, $L_v \gg L_{\omega}$ because the velocity is dominated by the large length scales and the vorticity by the small length scales.  However, zombie turbulence does not have a Kolmogorov spectrum and is dominated by large vortices. In flows dominated by large vortices,  
$L_v$ and $L_{\omega}$ are both approximately the mean size of the vortices.  Therefore, in a flow dominated by zombie vortices, 
\begin{equation}
M\!a_{rms} \simeq (\Omega_0 L_{zv} /c_s) \, Ro_{rms}, \label{zdominated}
\end{equation}
where $L_{zv}$ is the characteristic diameter of a zombie vortex. In a flow with constant vertical gravity (either an anelastic or a fully compressible flow with $\gamma =5/3$), Eq.~(\ref{home}) determines $c_s$, and Eq.~(\ref{zdominated}) then shows that a flow with zombie vortices has   
\begin{equation}
M\!a_{rms} \simeq \sqrt{2/5}  \,\, Ro_{rms} \, \beta^{-1} \, (L_{zv}/H_0). \label{linear}
\end{equation}
In all of the calculations presented in this paper, $L_{zv} \simeq \Delta$, so using Eq.~(\ref{delta}) for $\Delta$,  Eq.~(\ref{linear}) becomes
\begin{equation}
M\!a_{rms} \simeq  {\frac{\sqrt{2/5}}{3 \pi}} \, {\frac{L_y}{H_0}} \, Ro_{rms} \, \simeq   0.067 \,\, Ro_{rms}, \label{almost}
\end{equation}
where we used the fact that in our computations $L_y = H_0$. Equation~(\ref{almost}) shows that $M\!a_{rms}$ depends on the size of the computational box,  is consistent with  Figs.~\ref{fig:exam1} --~\ref{fig:exam3} at late times, and explains why our calculations produce zombie turbulence with $M\!a_{rms} \simeq 10^{-2}$.\footnote{However, note that in Figs.~\ref{fig:exam2} and~\ref{fig:exam3}, the values of $Ro_{rms}(t)$ have plateaued at late times, but the values of $M\!a_{rms}(t)$ are still slowly growing. This suggests that these flows have not quite yet reached a statistically steady state. The only way in which $M\!a_{rms}(t)$ can grow while keeping $Ro_{rms}(t)$ fixed is if $V_{eddy}(L_v)/V_{eddy}(L_{\omega})$ is still growing, and the latter is indicative that the reverse cascade of energy is still continuing at late times in the flows in Figs.~\ref{fig:exam2} and~\ref{fig:exam3}.} 

Equations~(\ref{linear}) and (\ref{almost}) do not apply to a PPD for two reasons. First, due to the fact that the vertical component of the protostellar gravity is linear in $z$, rather than constant, the sound speed is given by Eq.~(\ref{linear_g}b) rather than Eq.~(\ref{home}). Therefore, Eq.~(\ref{linear} is replaced by
\begin{equation}
M\!a_{rms} \simeq (L_{zv}/H) \,\, Ro_{rms}. \label{PPDMA}
\end{equation}
The second reason that Eq.~(\ref{almost}) does not apply to a PPD is we expect $L_{zv} \simeq H$ for zombie vortices in a PPD, rather than $L_{zv} \simeq \Delta$. This expectation follows from the observation that in our numerical simulations, zombie vortices grow until they run into each other in the $x$ direction (or they run into the $x$-boundary of the computational domain).  Because the spacing of the dominant critical layers in $x$ is $\Delta$, the spacing of the initial vortices is also $\Delta$ and the vortices run into each other and stop growing when $L_{zv} \simeq \Delta$. (We explicitly chose the parameters in our studies to have $L_x > \Delta$ so that $L_{zv}$ would be determined by $\Delta$ and not by $L_x$.) However in a fully compressible flow, it is well known that vortex diameters are often constrained by the fact that the vortices must be subsonic, otherwise the vortices rapidly dissipate their energy via acoustic radiation. In particular, the change in the velocity from one side of the vortex to the opposite side must be less than the sound speed. Therefore, in fully compressible flows, vortices stop growing when they become large enough to be supersonic, which can occur while the vortices are still too small to run into each other.   By definition of the Rossby number, the characteristic relative velocity of a zombie vortex is  $|Ro_{rms} \, \Omega_0 \, L_{zv}|$. However, the characteristic change in the velocity of the total flow across the $x$-diameter of the vortex is $\sim$~$|\Omega_0 L_{zv}|$ because the vortex is embedded in the Keplerian disk with a shear of $|q\Omega_0|$.  For the velocity change across the diameter of the vortex to be less the sound speed,  $L_{zv}$ must be less than $\sim$~$H$. In a PPD,  $H \ll \Delta$, so we would expect that the diameters of zombie vortices  have their growth limited by the constraint that the vortices remain subsonic, and not by the size $\Delta$.  Using $L_{zv} \simeq H$ in Eq.~(\ref{PPDMA}) makes the  Mach numbers of the zombie vortices in a PPD of order $|Ro_{rms}|$. Note that in some fully compressible flows that it is possible that $H$ is greater than $\Delta$. In fact, in $\S$\ref{subsec:reverse} we showed that for the fully compressible flow examined in this paper that $H_0 \simeq 5 \Delta$. Therefore for the fully compressible flow examined in this paper, we expect, and have observed numerically, that $L_{zv} \simeq \Delta$, that vortex diameters are not restricted by the subsonic constraint, that the zombie vortices remain subsonic, and that Eq.~(\ref{almost}) is valid.

Figs.~\ref{fig:XZ}d, \ref{fig:XY}d, and~\ref{fig:athena} show that at late times zombie turbulence fills the computational domain. We can quantify this space-filling property by defining a spatial filling factor for the turbulent vorticity: $f_{\!Ro}(\eta, t)$ is  the volume fraction of the computational domain that has $|Ro(x,y,z,t)| \ge \eta$.   Fig.~\ref{fig:fill} shows that for the anelastic flow in Fig.~\ref{fig:exam2}, approximately $10\%$ of the flow is filled with vortices with Rossby numbers with magnitudes greater than 0.3; $30\%$ with magnitudes greater than 0.2; and almost $60\%$ with magnitudes greater than 0.1. These fill factors may still be growing in time.  The fill factors in Fig.~\ref{fig:fill} are representative of all of our anelastic calculations. For the flow in our fully compressible {\it Athena} simulations, the fill factors are slightly larger with approximately $20\%$  of the flow is filled with vortices with Rossby numbers with magnitudes greater than 0.3; $45\%$ with magnitudes greater than 0.2; and almost $75\%$ with magnitudes greater than 0.1.

\unitlength1mm{
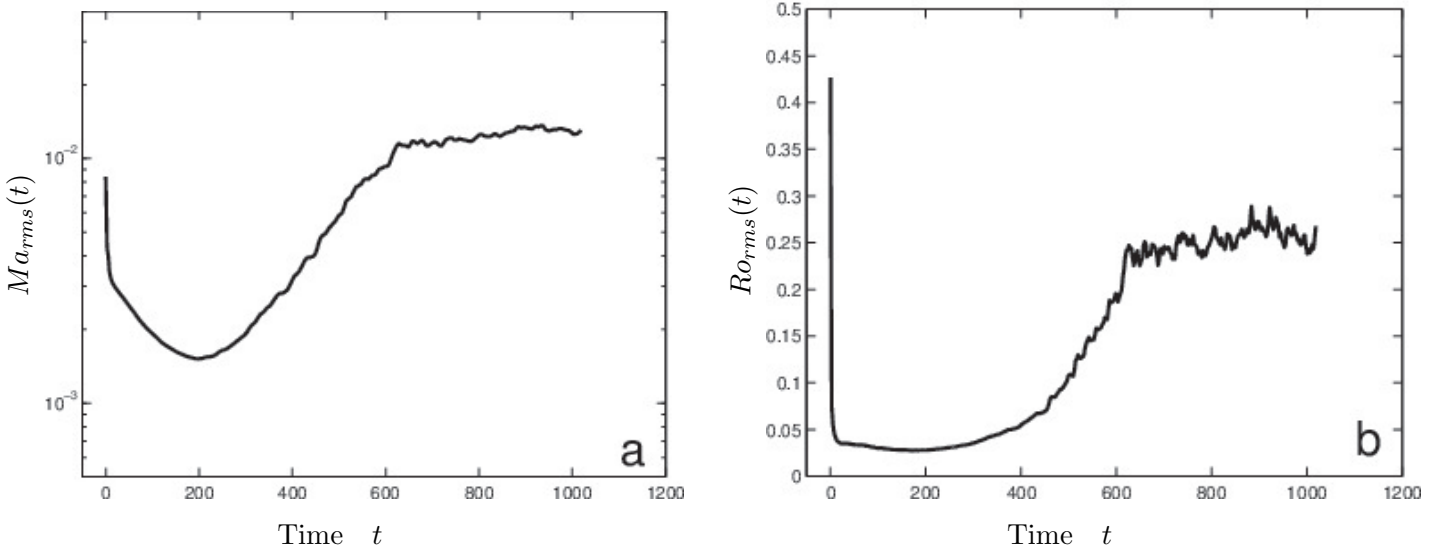
\begin{figure}[H]
\begin{center}
\begin{picture}(0,76)
\put(-90,6){\epsfig{figure=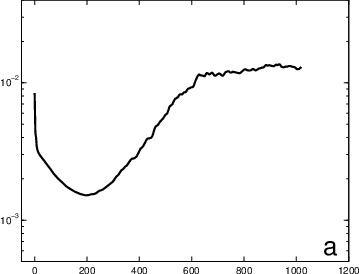,width=85mm}}   
\put(6, 6){\epsfig{figure=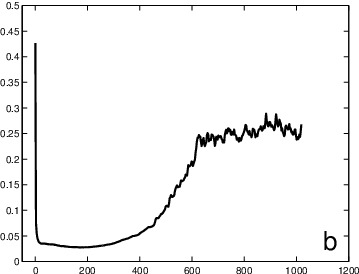,width=87mm}} 
\put(-95,33) {\rotatebox{90}{$Ma_{rms}(t)$}}
\put(-59,0) {Time \,\, $t$}
\put(1,33) {\rotatebox{90}{$Ro_{rms}(t)$}}
\put(38,0) {Time \,\, $t$}
\end{picture}
\renewcommand{\baselinestretch}{1.0}
\caption{\label{fig:exam1} 
For the same anelastic flow as in Figs.~\ref{fig:XZ} and~\ref{fig:XY}, the time evolution of (a) rms Mach number $M\!a_{rms}(t)$, and (b)  rms Rossby $Ro_{rms}(t)$.  The rms Mach and Rossby numbers both rapidly plummet due to hyperviscosity but grow after the zombie vortex instability sets in and eventually reach a plateau.  All of our calculations with zombie turbulence show late-time values of $Ro_{rms}(t)$ between $0.2$ and $0.3$.
}   
\renewcommand{\baselinestretch}{2.0}
\end{center}
\end{figure}}

\unitlength1mm{
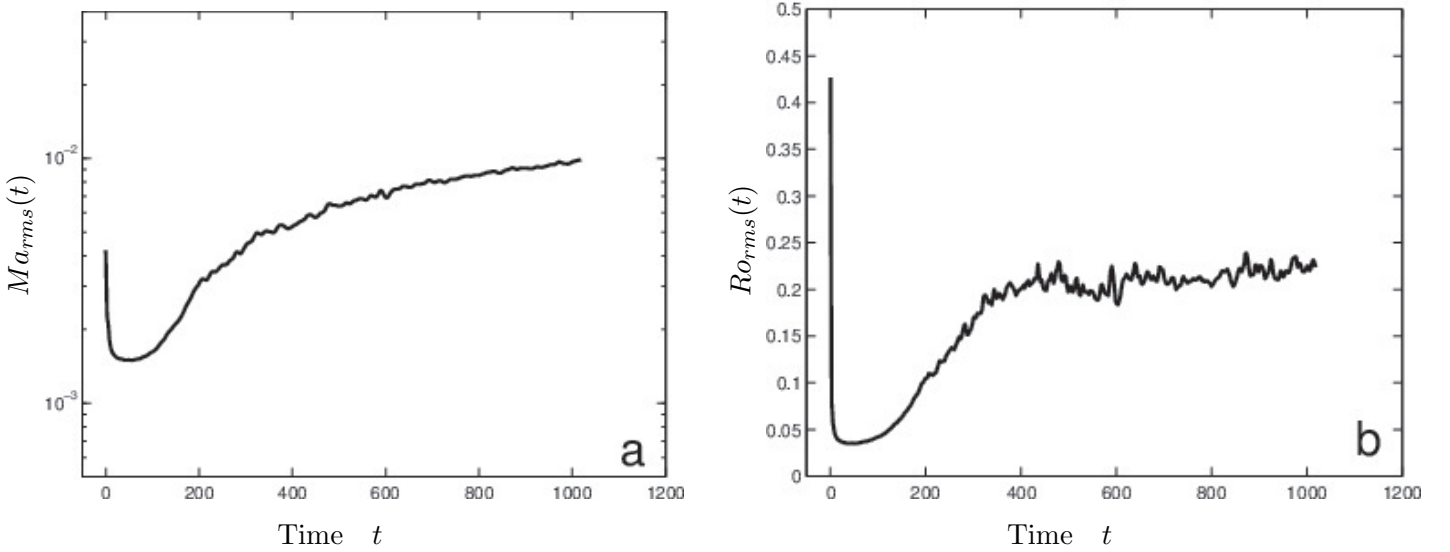
\begin{figure}[h]
\begin{center}
\begin{picture}(0,76)
\put(-90,6){\epsfig{figure=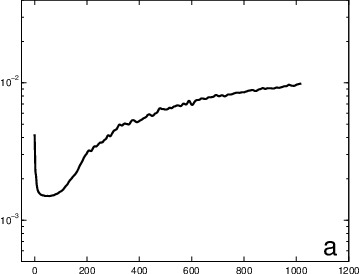,width=85mm}}   
\put(6, 6){\epsfig{figure=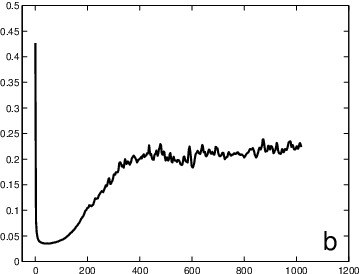,width=87mm}} 
\put(-95,33) {\rotatebox{90}{$Ma_{rms}(t)$}}
\put(38,0) {Time \,\, $t$}
\put(-59,0) {Time \,\, $t$}
\put(1,33) {\rotatebox{90}{$Ro_{rms}(t)$}}
\end{picture}
\renewcommand{\baselinestretch}{1.0}
\caption{\label{fig:exam2} 
Time evolution of $M\!a_{rms}$ and $Ro_{rms}$ as plotted in Fig.~\ref{fig:exam1}, but for the anelastic flow corresponding to the solid curve in Fig~\ref{fig:growenergy}a. This flow has the same initial amplitude of the Kolmogorov noise as the flow in Fig.~\ref{fig:exam1}, but has $\beta=2$ rather than $1$.}
\renewcommand{\baselinestretch}{2.0}
\end{center}
\end{figure}}

\unitlength1mm{
\begin{figure}[H]
\begin{center}
\begin{picture}(0,70)
\put(-90,6){\epsfig{figure=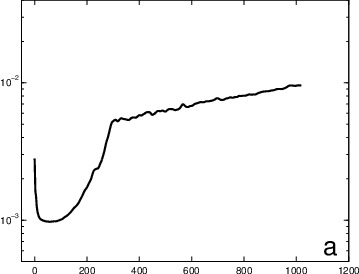,width=85mm}} 
\put(6, 6){\epsfig{figure=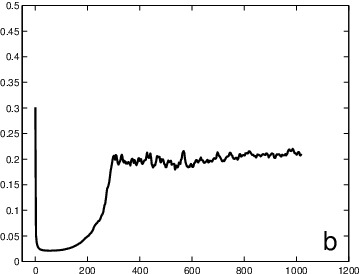,width=87mm}} 
\put(-95,33) {\rotatebox{90}{$Ma_{rms}(t)$}}
\put(38,0) {Time \,\, $t$}
\put(-59,0) {Time \,\, $t$}
\put(1,33) {\rotatebox{90}{$Ro_{rms}(t)$}}
\end{picture}
\renewcommand{\baselinestretch}{1.0}
\caption{\label{fig:exam3} 
Time evolution of $M\!a_{rms}$ and $Ro_{rms}$ as in Fig.~\ref{fig:exam2}. The flow parameters are the same as in Fig.~\ref{fig:exam2}, but here the rms Mach number of the initial Kolmogorov noise is two-thirds the value in Fig.~\ref{fig:exam2}. 
After $t \simeq 500$~yrs, many of the statistical properties of the flows in Fig.~\ref{fig:exam2} and~Fig.~\ref{fig:exam3} are nearly the same, which shows that the flows are being drawn to an attractor that is independent of the details of the initial conditions.}   
\renewcommand{\baselinestretch}{2.0}
\end{center}
\end{figure}
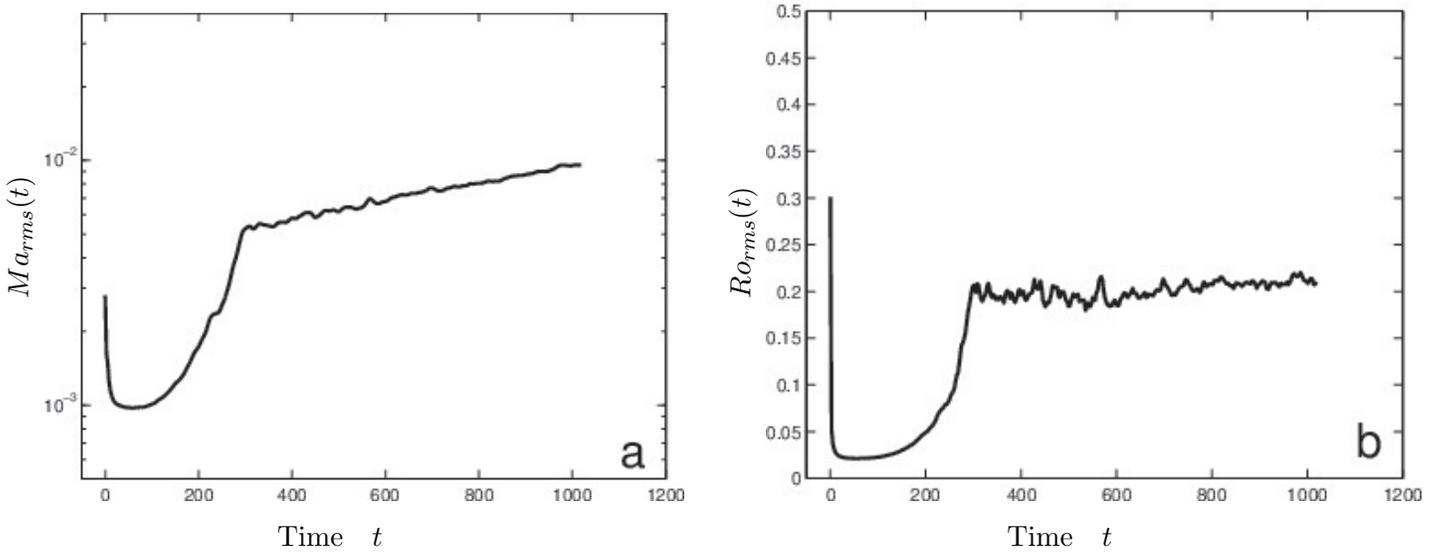}

\unitlength1mm{
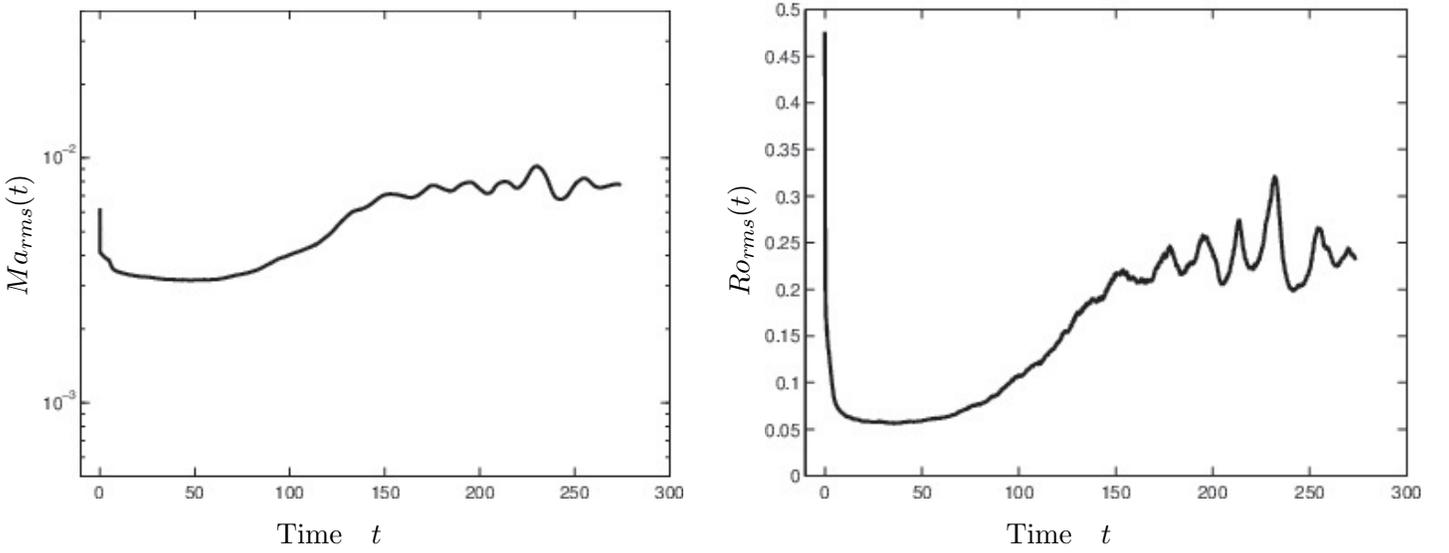
\begin{figure}[H]
\begin{center}
\begin{picture}(0,70)
\put(-90,6){\epsfig{figure=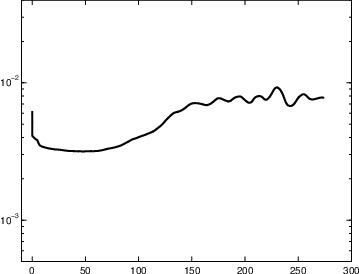,width=85mm}}  
\put(6, 6){\epsfig{figure=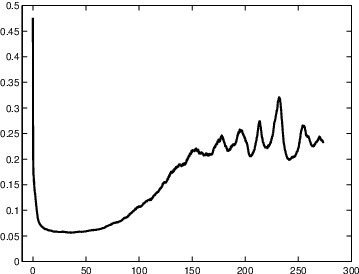,width=87mm}} 
\put(-95,33) {\rotatebox{90}{$Ma_{rms}(t)$}}
\put(38,0) {Time \,\, $t$}
\put(-59,0) {Time \,\, $t$}
\put(1,33) {\rotatebox{90}{$Ro_{rms}(t)$}}
\end{picture}
\renewcommand{\baselinestretch}{1.0}
\caption{\label{fig:exam4} 
Time evolution of $M\!a_{rms}$ and $Ro_{rms}$ for the compressible flow shown in Figs.~\ref{fig:growenergy}b and~\ref{fig:athena}
computed with the Godunov finite-volume code {\it Athena}.  The late-time values of $Ro_{rms}$ of this compressible flow with $\beta=2$ 
are similar to the values found in the anelastic flows computed with a pseudo-spectral code with $\beta=2$ as shown in  Figs.~\ref{fig:exam2}b and~\ref{fig:exam3}b.
}   
\renewcommand{\baselinestretch}{2.0}
\end{center}
\end{figure}}

\unitlength1mm{
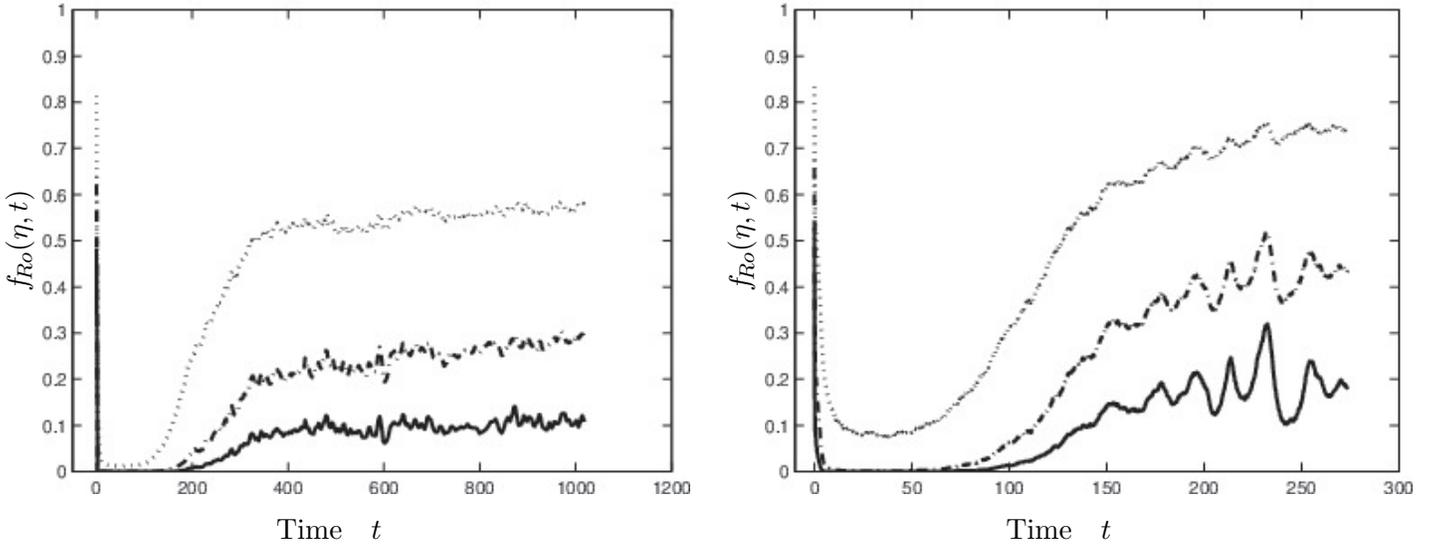
\begin{figure}[H]
\begin{center}
\begin{picture}(0,70)
\put(-90,6){\epsfig{figure=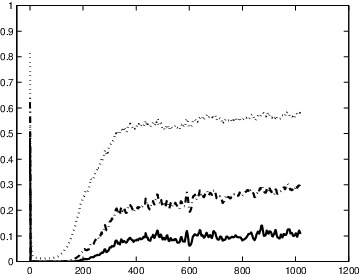,width=86mm}} 
\put(6, 6){\epsfig{figure=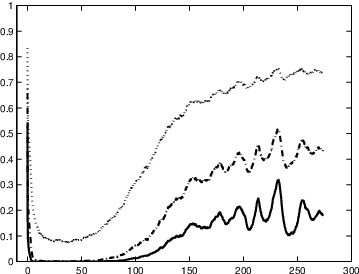,width=86mm}} 
\put(-95,33) {\rotatebox{90}{$f_{\!Ro}(\eta, t)$}}
\put(38,0) {Time \,\, $t$}
\put(-59,0) {Time \,\, $t$}
\put(1,33) {\rotatebox{90}{$f_{\!Ro}(\eta, t)$}}
\end{picture}
\renewcommand{\baselinestretch}{1.0}
\caption{\label{fig:fill} 
Time evolution of the spatial filling factor $f_{\!Ro}(\eta, t)$.   Dotted line for $\eta =0.1$; dot-dashed line for  $\eta =0.2$; solid line for $\eta =0.3$.  
(a) Left panel: The anelastic flow as in Figs.~\ref{fig:exam2} and~\ref{fig:growenergy}a (solid curve).  These filling factors are typical of all of our anelastic calculations.
(b) Right panel: The fully compressible flow as in Figs.~\ref{fig:growenergy}b, ~\ref{fig:athena}, and~\ref{fig:exam4}.}
\renewcommand{\baselinestretch}{2.0}
\end{center}
\end{figure}}

\section{DISCUSSION} \label{sec:discussion}

\subsection{Summary} \label{subsec:summary}

We have shown that Keplerian disks are unstable to the purely hydrodynamic ZVI instability when a vertical component of gravity from the central star or protostar is included and the vertical thermodynamic structure is stably stratified.  It may seem counterintuitive that introducing stable stratification makes a flow more unstable.  Yet we have demonstrated that vertical stratification introduces baroclinic critical layers into the flow.  In this work, we have explored ZVI in the limit of infinite cooling time.  ZVI has a fast growth rate, requiring only  a few hundred local ``years'' to produce large-magnitude turbulence.  Even if the instability is initially  spatially confined to a small region, the entire domain fills rapidly with turbulence with rms Rossby numbers between 0.2 and 0.3.   We have argued that in a fully compressible PPD, we expect that the Mach number of the space-filling zombie turbulence will be approximately equal to its Rossby number.  When turbulence created by ZVI reaches a statistically-steady state, it is neither homogeneous nor isotropic, and the pattern of the turbulence is highly asymmetric with respect to cyclonic and anticyclonic vorticity. The flow is dominated by  large, elongated, anticyclonic vortices and narrow cyclonic layers of vorticity that are aligned in the stream-wise direction (\ie the azimuthal direction in a PPD). This pattern makes zombie turbulence easily identifiable and distinct from other forms of turbulence. In a PPD, the vortex size and the spacing between the layers would be expected to be of order the vertical pressure scale height. Our calculations show that zombie turbulence is robust and does not decay. In MPJH13, the permanence of the turbulence was shown to be due to the fact that it draws its energy from the energy stored in the Keplerian shear.  Disks filled with  statistically-steady zombie turbulence appear to be attracting solutions that are independent of the details of the  initial conditions that trigger the turbulence; properties,  such as the energy of the non-Keplerian velocity, the energy spectrum as a function of wavenumber,  and the probability distributions of the Mach and Rossby numbers are independent of the triggering perturbations.

\subsection{Why ZVI Was Not Seen in Previous Studies} \label{subsec:why}

The robustness of ZVI prompts the question of how it was missed in previous studies.  Part of the answer is that ZVI has four necessary ingredients: rotation, a component of gravity along the (vertical) rotation axis, horizontal shear, and stable vertical stratification.  These ingredients are all present in a PPD, but many previous studies neglected stable vertical stratification. Any stability study that does not include all four ingredients -- whether a numerical calculation or a laboratory experiment that uses a constant-density fluid -- cannot produce ZVI. 

However, there are several others reasons that previous studies did not find ZVI.

\begin{itemize}
\item Sufficient numerical resolution and/or control over numerical dissipation is required to resolve critical layers. Figs.~\ref{fig:XZ} --~\ref{fig:athena} show that these layers are very thin.   Simulations using the Godunov finite-volume code {\it Athena} show that ZVI is present only when more than 128  grid points per vertical pressure scale height are used in the radial direction.  Thus the instability could not be observed in the simulations of vertically stratified disks by \citet{FP06} who used 30 radial points per pressure scale height or by  \citet{FS03,OML09} who used 64 or fewer.

\item Sufficient integration time is needed for vorticity at small length scales to reverse cascade to larger scales, for the cyclonic vorticity to homogenize, and for the anticyclones to merge into vortices that are sufficiently large and spatially scattered  (as in Fig~\ref{F:zombie_cartoon}b) to excite critical layers. It then takes additional time for dipolar vortex layers to form and several more vortex turn-around times for them to roll up and create the  first generation of zombie vortices.  In our studies (\cf Fig.~\ref{fig:growenergy}), it takes at least 50 years for the zombie vortices to fill the domain, for energy to be drawn from the  Keplerian shear by the vortices, and for the non-Keplerian energy component to increase substantially.  If a calculation were terminated early (\eg after 8 years, as in Fig.~\ref{fig:balbus}), then ZVI would not be observed.
\end{itemize}

\subsection{Outstanding Issues with ZVI}

In BM05, we hypothesized that vortices themselves could directly transport angular momentum radially outward in a PPD.  However, we found that the rate of transport, as parameterized by $\alpha \equiv \langle \rho v_x \, v_y \rangle/(c_s^2 \langle \rho \rangle)$, was a factor of $\sim$100 times smaller than the values needed for the observed rate of star formation from a protostar, where $\langle Q \rangle$ means the volume-averaged value of $Q$ over the entire domain. The value of $\alpha$ is of order:
\begin{equation}
\alpha\sim M\!a_{rms}^2 \,\, f_{\!M\!a} \,\, f_{\rho} \, c,
\end{equation}
where $c$ is the correlation between $v_x$ and $v_y$, $f_{\!M\!a}$ is the time-averaged spatial fill factor of the flow with $M\!a_{rms}$, and $f_{\rho}$ is the ratio of the Mach-number-weighted average density of the gas to the density of the gas in the mid-plane of the PPD. In BM05, the small values of $\alpha$ were due to the small values of the correlation $c$ because of the inherent symmetry of a vortex. Vortices in the disk are approximately mirror symmetric with respect to the $x$--$z$ plane that passes through the vortex center; thus, the values of $\langle \rho v_x \, v_y \rangle$ on either side of the symmetry plane nearly cancel. We therefore believe that zombie vortices cannot {\it directly} produce a sufficiently large  $\alpha$ for star formation. 

We now hypothesize that zombie turbulence in a fully compressible fluid {\it indirectly} may produce much larger values of $\alpha$.  A fully compressible fluid has three orthogonal velocity components \citep{chandra61}: two are rotational and divergence-free (\ie the poloidal and toroidal components, which anelastic flows have), and one is an irrotational component with a non-zero divergence (\ie the  curl-free  potential flow, which anelastic flows do not have).  We suspect that the rotational divergence-free zombie turbulence will excite acoustic modes (\ie equipartition of energy among the poloidal, toroidal and potential components). \citet{Johnson2005,Shen2006,lesur2010,lyra2011,raettig2013} have shown that acoustic waves in a PPD transport angular momentum radially outward with  values of $\alpha$ of order $M\!a_{rms}^2$.  Acoustic waves have correlations $c$ of order unity.  For acoustic waves launched by zombie vortices, we estimate $c \simeq 1$,  $M\!a_{rms} \simeq 0.2$, $f_{\!M\!a} \simeq 0.2$, and $f_{\rho} \simeq 0.3$, which yields a value of $\alpha \sim 2 \times 10^{-3}$, consistent with values inferred from observed rates of star formation.  Our estimate that the fill factor $f_{\rho}$ is small is based on the  numerical simulations by BM05 and \citet{lesur2009stability} that found that vortices can exist above and below the disk midplane, but the midplane itself at $|z| < 0.1 H$ is devoid of laminar vortices with $|Ro| > 0.1$. The calculation of $\alpha$ in a fully compressible calculation of a PPD will be the subject of future papers. However, in this paper we have shown, regardless of the value of $\alpha$, that ZVI destabilizes a PPD and creates space-filling turbulence with rms Rossby numbers between 0.2 and 0.3. 

Other outstanding issues include: (i) varying different profiles of stable vertical stratification, (ii) determining how close to the nearly unstratified midplane can turbulence from ZVI can penetrate, (iii) including explicit cooling terms, (iv) studying ZVI in global simulations, (v) exploring how ZVI interacts and competes with MRI, (vi) investigating how ZVI interacts and competes with streaming instabilities, and (vii) computing dust capture inside zombie vortices.

\acknowledgments

PSM is supported in part by NSF grants AST-0905801 and AST-1009907, and by NASA PATM grants NNX10AB93G and NNX13AG56G.  Part of the computational work used an allocation of computer resources from the Extreme Science and Engineering Discovery Environment (XSEDE), which was supported by National Science Foundation Grant No. OCI-1053575, and part was supported by NASA-HEC.  JAB is supported by NSF AST-1010052.  PH is supported by a Ziff Environmental Fellowship from Harvard University Center for the Environment.  DL is supported by a Hertz Foundation Fellowship, the National Science Foundation Graduate Research Fellowship under Grant No. DGE 1106400, a Kavli Institute for Theoretical Physics Graduate Student Fellowship, and partially by the Schneider Chair in Physics to Eliot Quataert.  PSM, JAB, PH and DL would like to thank the Kavli Institute for Theoretical Physics (KITP) for hosting us while drafting this paper (KITP is supported in part by the NSF grant PHY-1125915).  The authors thank DeJon Phillips for graphic design services.

\end{document}